\documentclass[fleqn,usenatbib]{mnras}

\usepackage{newtxtext,newtxmath}

\usepackage[T1]{fontenc}

\DeclareRobustCommand{\VAN}[3]{#2}
\let\VANthebibliography\thebibliography
\def\thebibliography{\DeclareRobustCommand{\VAN}[3]{##3}\VANthebibliography}

\usepackage{graphicx}	
\usepackage{amsmath}	
\usepackage{adjustbox}
\usepackage{tabularx}
\usepackage[figuresright]{rotating}

\def\app#1#2{%
  \mathrel{%
    \setbox0=\hbox{$#1\sim$}%
    \setbox2=\hbox{%
      \rlap{\hbox{$#1\propto$}}%
      \lower1.1\ht0\box0%
    }%
    \raise0.25\ht2\box2%
  }%
}
\def\approxprop{\mathpalette\app\relax}

\title[AT2024wpp Modelling]{Multiwavelength Modelling of the Luminous Fast Blue Optical Transient AT2024wpp}

\author[Omand et al.]{
Conor M. B. Omand$^{1}$\thanks{E-mail: c.m.omand@ljmu.ac.uk}\href{https://orcid.org/0000-0002-9646-8710}{\includegraphics[scale=0.5]{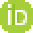}},
Nikhil Sarin$^{2,3}$\href{https://orcid.org/0000-0003-2700-1030}{\includegraphics[scale=0.5]{ORCIDiD_icon16x16.eps}}, 
Gavin P. Lamb$^{1}$\href{0000-0001-5169-4143}{\includegraphics[scale=0.5]{ORCIDiD_icon16x16.eps}},
Daniel A. Perley$^{1}$\href{https://orcid.org/0000-0001-8472-1996}{\includegraphics[scale=0.5]{ORCIDiD_icon16x16.eps}},
Andrew Mummery$^4$, 
\newauthor
Hamid Hamidani$^5$\href{https://orcid.org/0000-0003-2866-4522}{\includegraphics[scale=0.5]{ORCIDiD_icon16x16.eps}}, 
Steve Schulze$^6$\href{https://orcid.org/0000-0001-6797-1889}{\includegraphics[scale=0.5]{ORCIDiD_icon16x16.eps}}, 
Emma R. Beasor$^{1}$\href{https://orcid.org/0000-0003-4666-4606}{\includegraphics[scale=0.5]{ORCIDiD_icon16x16.eps}},
Aleksandra Bochenek$^{1}$\href{https://orcid.org/0009-0008-2714-2507}{\includegraphics[scale=0.5]{ORCIDiD_icon16x16.eps}},
\newauthor
Helena-Margaret S. Grabham$^1$,
Sorcha R. Kennelly$^1$\href{https://orcid.org/0009-0004-3767-2430} {\includegraphics[scale=0.5]{ORCIDiD_icon16x16.eps}},
Nguyen M. Khang$^{1}$\href{https://orcid.org/0000-0001-9657-8728}{\includegraphics[scale=0.5]{ORCIDiD_icon16x16.eps}},
Shiho Kobayashi$^{1}$\href{https://orcid.org/0000-0001-8172-4411}{\includegraphics[scale=0.5]{ORCIDiD_icon16x16.eps}},
\newauthor
Genevieve Schroeder$^7$\href{https://orcid.org/0000-0001-9915-8147}{\includegraphics[scale=0.5]{ORCIDiD_icon16x16.eps}},
William N. Stone$^{1}$\href{https://orcid.org/0009-0002-5946-8268} {\includegraphics[scale=0.5]{ORCIDiD_icon16x16.eps}},
Cairns Turnbull$^{1}$\href{https://orcid.org/0009-0002-7044-5358}{\includegraphics[scale=0.5]{ORCIDiD_icon16x16.eps}},
Jacob Wise$^{1}$\href{https://orcid.org/0000-0003-0733-2916}{\includegraphics[scale=0.5]{ORCIDiD_icon16x16.eps}}
\\
$^{1}$Astrophysics Research Institute, Liverpool John Moores University, Liverpool Science Park IC2, 146 Brownlow Hill, Liverpool, UK, L3 5R \\
$^{2}$Kavli Institute for Cosmology, University of Cambridge, Madingley Road, CB3 0HA, UK\\
$^{3}$Institute of Astronomy, University of Cambridge, Madingley Road, CB3 0HA, UK\\
$^4$School of Natural Sciences, Institute for Advanced Study, 1 Einstein Drive, Princeton, NJ 08540, USA \\
$^5$Astronomical Institute, Graduate School of Science, Tohoku University, Sendai 980-8578, Japan\\
$^6$Center for Interdisciplinary Exploration and Research in Astrophysics (CIERA), Northwestern University, 1800 Sherman Ave., Evanston, IL 60201, USA\\
$^7$Department of Astronomy, Cornell University, Ithaca, NY 14853, USA
}

\date{Accepted XXX. Received YYY; in original form ZZZ}

\pubyear{2025}

\begin{document}
\label{firstpage}
\pagerange{\pageref{firstpage}--\pageref{lastpage}}
\maketitle

\begin{abstract}
Luminous fast blue optical transients (LFBOTs) are a growing class of enigmatic energetic transients.  They show fast rises and declines, high temperatures throughout their evolution, and non-thermal emission in radio and X-rays.  Their power source is currently unknown, but proposed models include engine-driven supernovae, interaction-powered supernovae, shock cooling emission, intermediate mass black hole tidal disruption events (IMBH TDEs), and Wolf-Rayet/black hole mergers, among others.  AT2024wpp is the most optically luminous LFBOT to date and has been observed extensively at multiple wavelengths, including radio, optical, UV, and X-rays.  We take models from multiple scenarios and fit them to the AT2024wpp optical, radio, and X-ray light curves to determine which of these scenarios can best describe all aspects of the data.  We show that none of the multiwavelength light curve models can reasonably explain the data, and that other physical arguments disfavour models with homologously expanding ejecta.  We discuss how a stellar mass/IMBH TDE of a low mass star can be tested with late-time observations, and what other scenarios could possibly explain the broadband data.
\end{abstract}




\section{Introduction}

Recent high-cadence wide-field surveys, such as the Zwicky Transient Facility \citep[ZTF,][]{ztf_paper} and the Asteroid Terrestrial-impact Last Alert System \citep[ATLAS,][]{Tonry2011, Tonry2018}, have led to the discovery and characterization of several new types of fast optical transients \citep{Drout2014, Arcavi2016, Pursiainen2018, Ho2023cow}.  These transients evolve faster than typical supernovae (SNe) and can have peak luminosities similar to superluminous supernovae \citep[SLSNe, ][]{Gal-Yam2012, Nicholl2021, Gomez2024}.  These properties rule out typical $^{56}$Ni-powered core-collapse or thermonuclear SN models, since the mass of $^{56}$Ni required to explain the luminosity would be larger than the mass of the ejected material.  Further studies have shown that many of these transients are likely Type Ibn/Icn SNe \citep{Ho2023cow}; SNe powered by the interaction between the ejected material and circumstellar material (CSM).  However, a few of the more luminous fast optical transients have shown multi-component multiwavelength signatures consistent with a central engine \citep{Margutti2019, Ho2019, Ho2020, Ho2022, Metzger2022BHWR, Chrimes2024multi}; these transients have become known as luminous fast blue optical transients (LFBOTs).

The prototype LFBOT is AT2018cow \citep[`the Cow', ][]{Prentice2018}, the closest observed event so far ($z$ = 0.014).  
The optical emission rose to peak within $\sim$ 2.5 days, and peaked at an absolute magnitude $\sim -$20.5.  The optical spectra had only a few broad features, indicating expansion velocities $\gtrsim$ 0.1$c$ and temperatures $\sim$ 30 000 K.  Around 15 days, the spectrum began to show H and He features at velocities $\sim$ 3000 -- 4000 km s$^{-1}$, but showed no evidence of cooling to normal late-time SN temperatures \citep{Margutti2019, Perley2019, Xiang2021}.  The spectrum also showed narrow emission components ($v \lesssim 300$ km s$^{-1}$), consistent with CSM interaction \citep{Fox2019, Dessart2021}.  The Cow showed polarization of $\sim$ 7\% at 5.7d post-explosion, which quickly faded to show no polarization after 8 days except for a $\lesssim$ 1 day increase to $\sim$ 1.5\% in the blue at 13 days \citep{Smith2018, Maund2023}.  Follow-up observations also found a persistent UV source coincident with the Cow at $\sim$ years post-explosion \citep{Sun2022cow, Sun2023, Chen2023_1, Chen2023_2, Inkenhaag2023, Inkenhaag2025}.

The Cow was also bright at other wavelengths.  Soft X-ray emission was detected shortly after the explosion \citep{RiveraSandoval2018, Kuin2019}, and followed a shallow temporal decline until steepening at $\sim$ 20 days, when the optical lines at $\sim$ 3000 -- 4000 km s$^{-1}$ emerged \citep{Margutti2019}.  The soft X-ray light curve was highly variable, and quasi-periodic oscillations (QPOs) were claimed at 225 Hz \citep{Pasham2021} and 4 mHz \citep{Xiang2021}, giving constraints on the mass of the central object to be $<$ 850 $M_\odot$ and $10^3 - 10^5$ $M_\odot$, respectively.  The persistent source found at $\sim$ years post-explosion also showed soft X-ray emission \citep{Migliori2024}.  The Cow also showed bright hard X-ray, radio, and millimetre emission \citep{Ho2019, Margutti2019, Nayana2021}.  The radio evolved smoothly and showed a steep temporal drop at $\sim$ 1 month \citep{Ho2019}, indicating a different origin from the X-rays.  The atypical properties of the radio compared to radio SNe can be understood if the radiating electrons have a relativistic Maxwellian energy distribution \citep{Ho2022, Margalit2021, Margalit2024}.

The multiwavelength analysis from \citet{Margutti2019} found that the observed properties of the Cow can be explained by a central engine surrounded by an aspherical ejecta with a trans-relativistic polar component and a lower velocity equatorial ring or torus.  The central engine hypothesis for LFBOTs is also supported by the time variability of the X-rays, giant optical flares from AT2022tsd \citep{Ho2023tsd}, possible QPOs, and long-lasting UV/X-ray remnant.  The radio emission is likely caused by the interaction between the trans-relativistic component and surrounding high density ($10^5$ cm$^{-3}$), wind-like CSM \citep{Margutti2019, Ho2019} on scales of $\sim$ 10$^{16}$ cm.  For wind speeds of $\sim$ 1000 km s$^{-1}$ (typical of Wolf-Rayet stars), the mass loss rate is implied to be $\dot{M} = 10^{-4} - 10^{-3} M_\odot$ yr$^{-1}$ \citep{Margutti2019}, which is more than an order of magnitude higher than expected in Wolf-Rayet stars \citep{Barlow1981, Nugis2000, Adhyaqsa2020}.  This indicates either an extended CSM or a distinct physical origin \citep{Fox2019}.  In the early phase, the optical emission from the polar component is thought to arise from reprocessed X-ray emission from the central engine \citep{Piro2020, Uno2020fbot, Calderon2021, Chen2022, Uno2023}.  The photosphere from this component recedes into the equatorial torus as reprocessing becomes less efficient on the timescale where the lower velocity lines emerge.

Since the discovery of the Cow, a few other transients have been discovered with similar properties.  These include AT2018lug/ZTF18abvkwla \citep[`the Koala', ][]{Ho2020}, CSS161010 \citep{Coppejans2020, Gutierrez2024}, AT2020xnd/ZTF20acigmel \citep[`the Camel', ][]{Perley2021, Bright2022, Ho2022}, AT2020mrf \citep{Yao2022}, AT2022tsd \citep[`the Tasmanian Devil', ][]{Matthews2023, Ho2023tsd}, and AT2023fhn \citep[`the Finch', ][]{Chrimes2024fhn, Chrimes2024multi}.  These objects show some variety in peak luminosity and ejecta velocity, but all show hot, roughly featureless early spectra, optical luminosities similar to SLSNe, and bright X-ray and radio emission.  \citet{Somalwar2025} also discovered AT2024puz, which has similar properties to both LFBOTs and TDEs and evolves on a timescale intermediate between the two.  The LFBOT rate is estimated to be $<$0.1\% of the core-collapse SN rate \citep{Ho2023cow}, and they normally occur in metal-poor, low-mass starburst galaxies \citep{Ho2019, Michalowski2019, Coppejans2020, Lyman2020, Yao2022}.  These environments are similar to those of Type-I SLSN and long gamma-ray burst (GRB) progenitors \citep{Lunnan2014, Chen2015, Leloudas2015b, Angus2016, Chen2017, Schulze2018, Orum2020}, which may indicate that these transients all come from massive, low-metallicity progenitors.  

Although the ejecta structure of LFBOTs is somewhat understood, the nature of the progenitor that produces it is still unknown.  A few proposed scenarios include engine-driven SNe \citep[e.g.][]{Prentice2018, Liu2022, Omand2024}, failed SNe with prompt accretion disc formation \citep{Margutti2019, Perley2019, Quataert2019}, tidal disruption events (TDEs) from intermediate mass black holes \citep[IMBHs, ][]{Perley2019, Kuin2019, Gutierrez2024}, choked jets \citep{Gottlieb2022, Soker2022, Suzuki2022, Suzuki2024, Hamidani2025}, highly aspherical SNe \citep[`ellipsars', ][]{Dupont2022}, and the merger of a compact object and massive star \citep{Lyutikov2019, Uno2020fbot, Schroder2020, Metzger2022BHWR}.  CSM-interaction models have also been proposed \citep{Fox2019, Xiang2021, Pellegrino2022, Khatami2024}, but these are disfavoured due to the variability and long duration of the X-ray emission.  The central X-ray source could be either a magnetized nebula \citep[e.g. ][]{Omand2018, Vurm2021, Omand2025grb}, which can cause ejecta asymmetry due to Rayleigh-Taylor instabilities between the wind and ejecta \citep{Blondin2017, Suzuki2017, Suzuki2021, Omand2025Crab}; the inner region of a Super-Eddington accretion disc \citep{Sadowski2015}; or a choked jet \citep{Gottlieb2022}.  The transients can be infrared bright due to the dense outflow creating a light echo \citep{Metzger2023, Li2025} or from newly formed dust reprocessing emission from the central engine \citep{Omand2019}.

One of the newest additions to the LFBOT class is AT2024wpp \citep[`the Whippet'][]{Ho2024}.  AT2024wpp is one of the closest \citep[$z = 0.0868$,][]{Perley2024} and brightest ($L_{\rm bol, peak} > 10^{45}$ erg s$^{-1}$) LFBOTs to date, and the first one to be identified before the optical/UV peak \citep{Ho2024}.  A comprehensive multiwavelength dataset on the object has been collected, including NIR/optical/UV photometry and spectroscopy, optical polarization, radio and millimetre data, and X-ray data \citep{Srinivasaragavan2024wpp, Margutti2024, Schroeder2024, Pursiainen2025, Nayana2025, LeBaron2026, Perley2026}. AT2024wpp is featureless at early epochs, but shows weak H and He features after $\sim$ 30 days.  Late-time observations show no evidence for minute-timescale flares \citep{Ofek2025} similar to AT2022tsd \citep{Ho2023tsd}.  The optical peak is well described by a $\gtrsim$ 30 000 K blackbody, and the optical component remains well-described by a $\gtrsim$ 20 000 K blackbody out to late times, although there is some evidence for a power-law NIR excess after $\sim$ 20 days.  The X-ray emission hardens over time and shows a rebrightening at $\sim$ 50 days.  The radio emission rises quickly over $\sim$ 20-30 days before starting to decline.

This work attempts to find a physically plausible scenario for AT2024wpp that can explain both the thermal and non-thermal emission.  We approach this task by fitting several semi-analytic models to each component and using the results of those fits to determine whether different scenarios are viable.  We summarize the data we attempt to model in Section \ref{sec:data} and the models used in Section \ref{sec:mods}.  The results are shown in Section \ref{sec:res} and discussed in Section \ref{sec:disc}.  Finally, we conclude in Section \ref{sec:conc}.

\begin{table*}
\centering
\begin{tabular}{|ccc|}
\hline
Model & \textsc{Redback} Model Name & Reference \\
\hline
\multicolumn{3}{|c|}{\textit{Optical}} \\
\hline
Evolving Blackbody & evolving\_blackbody & \citet{Sarin_redback} \\
\hline
$^{56}$Ni-powered SN & arnett & \citet{Arnett1982}  \\
Magnetar-powered SN & general\_magnetar\_driven\_supernova & \citet{Sarin2022, Omand2024} \\
CSM-powered SN & csm\_nickel & \citet{Chatzopoulos2013, Villar2017, Jiang2020} \\
Fallback-powered SN & sn\_nickel\_fallback & \citet{Guillochon2018} \\
SN + Envelope Shock Cooling & shockcooling\_sapirwaxman\_and\_arnett & \citet{Sapir2017} \\
SN + CSM Shock Cooling & shock\_cooling\_and\_arnett & \citet{Piro2021} \\
Cocoon + Envelope Shock Cooling & shocked\_cocoon\_and\_arnett & \citet{Piro2018} \\
Cocoon + CSM Shock Cooling & shocked\_cocoon\_csm\_and\_arnett & \citet{Hamidani2025b,Hamidani2025} \\
Fallback TDE & tde\_fallback & \citet{Guillochon2013, Guillochon2018, Mockler2019} \\
Stream-Stream Collision TDE & stream\_stream\_tde & \citet{Piran2015, Ryu2020} \\
Cooling Envelope TDE & gaussianrise\_cooling\_envelope & \citet{Metzger2022CE, Sarin2024CE} \\
BH-WR Merger & wr-bh\_merger & \citet{Metzger2022BHWR} \\
\hline
\multicolumn{3}{|c|}{\textit{Radio and X-ray}} \\
\hline
Top-hat Jet & tophat\_redback & \citet{Lamb2018_late} \\
Gaussian Jet & gaussian\_redback & \citet{Lamb2018_late} \\
Power-Law Synchrotron & synchrotron\_pldensity & \citet{Rosswog2007, Chevalier2017} \\
Thermal Synchrotron & thermal\_synchrotron\_v2\_fluxdensity & \citet{Margalit2021, Margalit2024} \\
Accretion Disc & fitted & \citet{Mummery2020, Mummery2025, Mummery2024scaling} \\
Top-hat Jet + Accretion Disc & tophat\_redback \& fitted &  \\
\hline
\end{tabular}
\caption{The optical, radio, and X-ray models used to fit the data for AT2024wpp, and their associated references.}  
\label{tbl:models}
\end{table*}

\section{Data Modeled} \label{sec:data}

The data we model include the NIR/optical/UV photometry, radio data, and X-ray data presented in \citet{Perley2026}.  The photometry was gathered using a number of surveys and facilities, including the Zwicky Transient Facility \citep[ZTF;][]{Bellm2019, Graham2019, Dekany2020} using the Samuel Oschin 48-inch Schmidt telescope, IO:O and the Liverpool Infra-Red Imaging Camera (LIRIC) on the Liverpool Telescope, the Rainbow Palomar 60-inch telescope at Palomar Observatory, the Panoramic Survey Telescope and Rapid Response System (Pan-STARRS), Goodman High-Throughput Spectrograph (GHTS) on the Southern Astrophysical Research Telescope (SOAR), the ESO Faint Object Spectrograph and Camera v2 (EFOSC2) on the New Technology Telescope (NTT), the Focal Reducer/Low Dispersion Spectrograph 2 (FORS2) on the Very Large Telescope (VLT), the Ultraviolet-Optical Telescope (UVOT) on the Neil Gehrels \textit{Swift} Observatory, and the Wide Field Camera 3 (WFC3) on the Hubble Space Telescope (HST).  X-ray observations were taken by the X-Ray Telescope (XRT) on \textit{Swift}, and the radio observations were taken by the Karl G. Jansky Very Large Array (VLA) and Atacama Large Millimetre/submillimetre Array (ALMA).  Overall, the UVOIR photometric data has detections spanning $\sim$ 120 days and upper limits at $\sim$ 150 -- 220 days after the last detection.  The radio data spans the 3.0 -- 350 GHz frequency range and has detections up to $\sim$ 200 days post-discovery.  The X-ray data is only in one band (0.2 -- 10 keV), and lasts $\sim$ 80 days.  Upper limits were not included in model fits in either the optical or non-thermal models, although late-time upper limits are used as constraints in Section \ref{sec:disc}.

\section{Models and Fitting} \label{sec:mods}

\begin{figure}
    \centering
    \includegraphics[width=0.95\linewidth]{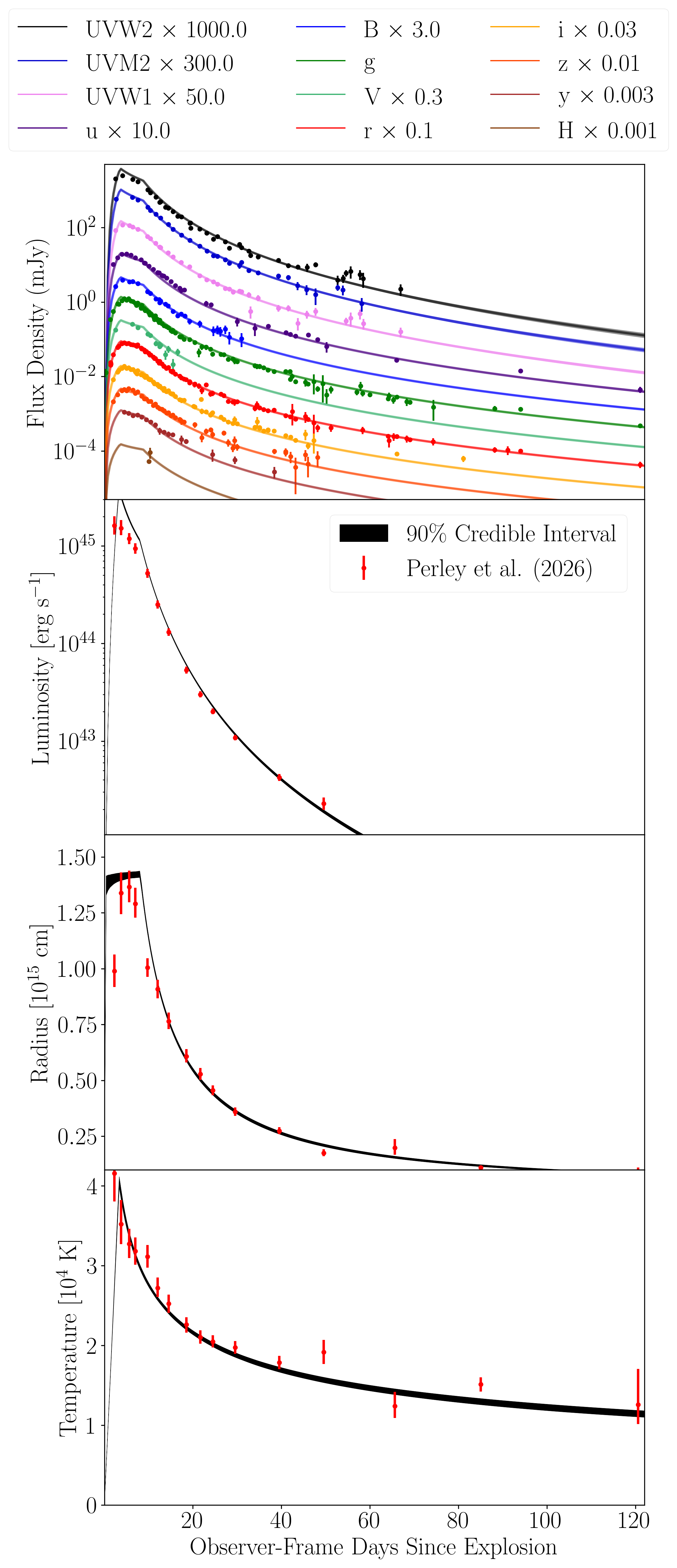}
    \caption{The multiband fit from the evolving blackbody (top) and the luminosity (second), radius (third), and temperature (bottom) derived from it. The shaded region shows the 90$\%$ credible interval. Values from \citet{Perley2026} are shown in red.}
    \label{fig:bbprops}
\end{figure}

We test various models by performing Bayesian inference to fit them to the multiband optical photometry and radio data using \textsc{Redback}\footnote{Version 1.15.1} \citep{Sarin_redback}.  We use the \textsc{PyMultiNest} sampler \citep{Buchner2014} implemented in \textsc{bilby} \citep{Ashton2019}, and sample in flux density with a Gaussian likelihood.  The priors used are listed in Appendix \ref{sec:priors}.  
We sample the initial time with a uniform prior of up to 20 days before the first observation, and include a host extinction term $A_V$ with a uniform prior between 0 and 2 mag for the optical data.  
A list of the models used is given in Table \ref{tbl:models}.

The first model used is an evolving blackbody model, which fits the transient by assuming the photospheric temperature and radius both have power-law rises and declines in time.  The temperature and radius peaks are not assumed to be at the same time.  While this will not reveal the physics of the LFBOT, it will allow us to characterize the photospheric radius and temperature of the system, which can provide insight into the properties that the more physical models need to have in order to describe the data.

The physical semi-analytic optical models we test can be divided into a few broad classes: SNe, shock cooling, TDEs, and stellar mergers.  These models were chosen to represent the wide range of phenomena that could potentially explain LFBOTs.  These models all calculate the bolometric luminosity and photospheric radius of one or more components, and convert that to observed emission by assuming each component is described by a blackbody spectral energy distribution.

\subsection{Supernova Models}

We test four SN models, which differ depending on their power source.  One is the standard $^{56}$Ni-powered SN, known as the Arnett model \citep{Colgate1969, Arnett1982}.  This model usually has difficulty explaining transients with high luminosities and low diffusion times, due to the constraint that the $^{56}$Ni mass must be lower than the ejecta mass.  The other SNe models also have $^{56}$Ni, but are primarily powered by a different source.  One model is the millisecond magnetar model \citep{Omand2024}, which was shown in \citet{Omand2024} to be a good fit to the LFBOT AT2020xnd \citep[`the Camel', ][]{Perley2021}.  Another is the CSM interaction model \citep{Chatzopoulos2013, Villar2017, Jiang2020}, powered by the conversion of kinetic energy into luminosity via the shock formed between the ejecta and CSM.  The final one is the fallback SN model \citep{Guillochon2018}, which is powered by the accretion of material onto a central black hole remnant.

\subsection{Shock Cooling Models}

Shock cooling emission is the radiation from shock heated material as it cools and expands after shock breakout \citep[e.g.][]{Margalit2022sbo}.  The shock that heats the material can come from either an SN explosion or a GRB jet, leaving a region of shocked material called the cocoon.  The material can either be the envelope of the star or an extended CSM.  This leads us to examine four shock cooling scenarios: the SN + envelope \citep{Sapir2017}, SN + extended CSM \citep{Piro2021}, cocoon + envelope \citep{Piro2020}, and cocoon + extended CSM \citep{Hamidani2025b,Hamidani2025}.  Shock cooling can emit on short timescales but cannot produce emission at later times, so we test all the shock cooling models alongside a $^{56}$Ni component to better explain late-time emission.

\subsection{Tidal Disruption Event Models}

We test three models for TDE emission, which vary how the optical emission is produced after the star is disrupted.  None of the models self-consistently solve the early pre-circularization emission, and instead they all assume a constant-temperature Gaussian rise until peak.  The first model assumes the emission is directly powered by fallback accretion onto the central black hole \citep{Guillochon2013, Guillochon2018, Mockler2019}.  The second model assumes the emission is powered by shocks resulting from the collisions of debris streams \citep{Piran2015, Ryu2020}.  The final TDE model assumes the bound debris forms a quasi-steady envelope which cools radiatively as it undergoes Kelvin-Helmholtz contraction \citep{Metzger2022CE, Sarin2024CE}.  For these models, we alter the default prior such that the lower bound of the black hole mass goes down to 1 $M_\odot$, since the TDE rise time is predominately set by the black hole mass and a timescale of $\sim$ 5 days requires an intermediate or stellar mass black hole.

\subsection{Stellar Merger Model}

The final optical model is the disruption and hyper-accretion of a Wolf-Rayet star onto a black hole companion, with a delay between the common envelope phase and the merger \citep{Metzger2022BHWR}.  The optical emission in this model is primarily powered by reprocessed X-rays from the inner accretion disc, although interaction between ejecta and CSM can also contribute at later times.

\subsection{Radio and X-ray Models}

\begin{figure*}
\includegraphics[width=\linewidth]{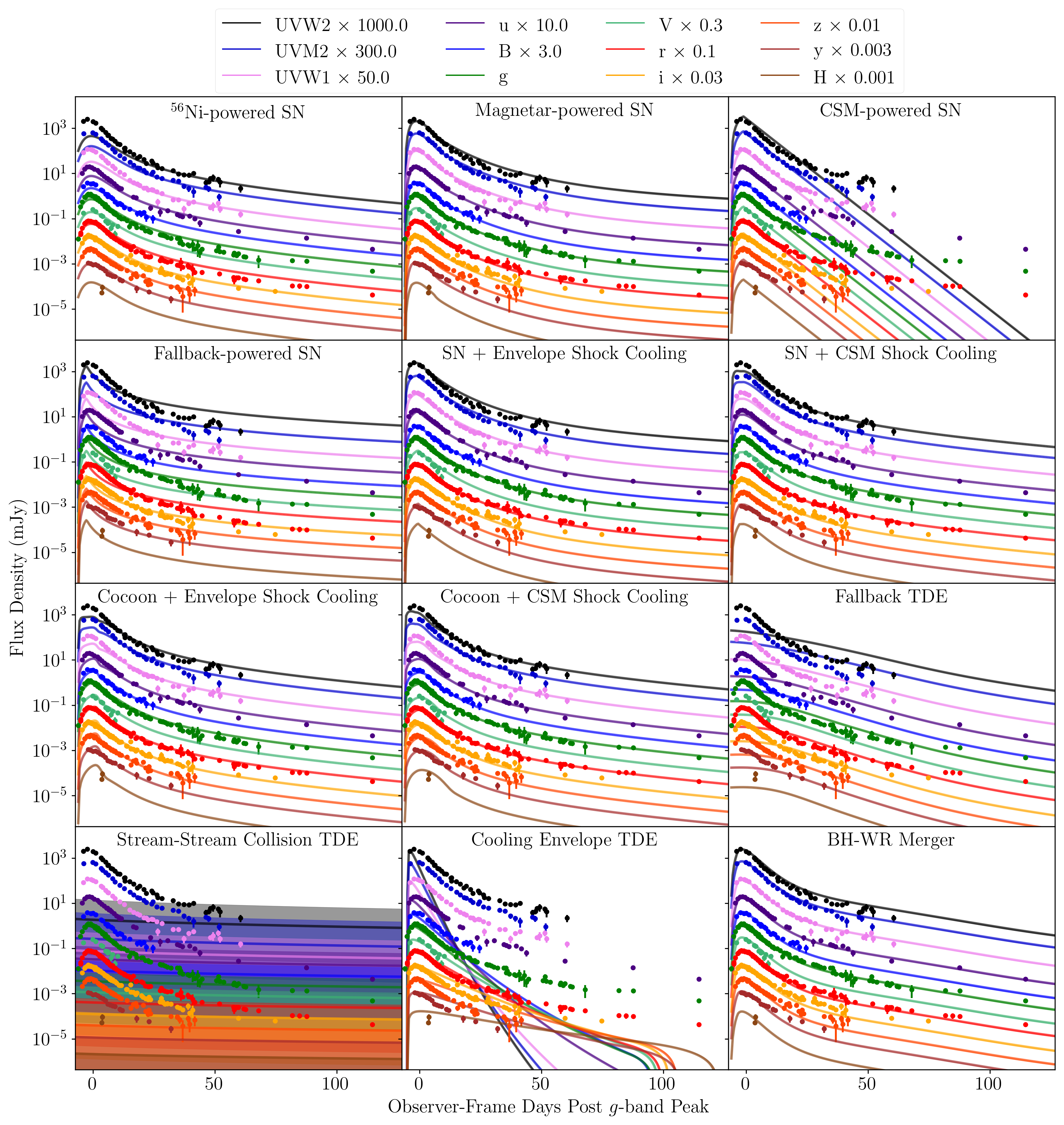}
\caption{The light curve fits to the broadband optical data for each model discussed in Section \ref{sec:mods} (see Table \ref{tbl:models}) using default priors.  The solid line shows the model with the highest likelihood while the shaded region shows the 90$\%$ confidence interval.} %
\label{fig:optfits_tfree}
\end{figure*}

To try and explain the radio and X-ray emission, we test models from GRB afterglows, CSM interaction from non-relativistic or trans-relativistic spherical blast waves, and from an accretion disc.  Since previous LFBOTs have shown evidence for multiple components in their non-thermal emission, the radio and X-ray data are fit separately.  We test two jet structures for the afterglow: a top-hat jet and a Gaussian jet \citep{Lamb2018_late}.  For CSM interaction, we test a power-law synchrotron model \citep{Rosswog2007, Chevalier2017}, where the emitting electrons have a power-law energy distribution, and a thermal synchrotron model \citep{Margalit2021, Margalit2024}, where the emitting electrons have a relativistic Maxwellian distribution.  The GRB afterglow models are assumed to be in a constant density medium since they decelerate further away from the progenitor, while the interaction models are assumed to be in a medium with a power-law density profile $\rho \propto r^{-\alpha}$, with $\alpha$ between 0 and 4.  The emission from the accretion disc is fit using models from \textsc{FitTeD} \citep{Mummery2025} imported into \textsc{Redback}, and we also examine a case with a jet and accretion disc.

\section{Results} \label{sec:res}

\subsection{Optical}

\begin{figure*}
\includegraphics[width=\linewidth]{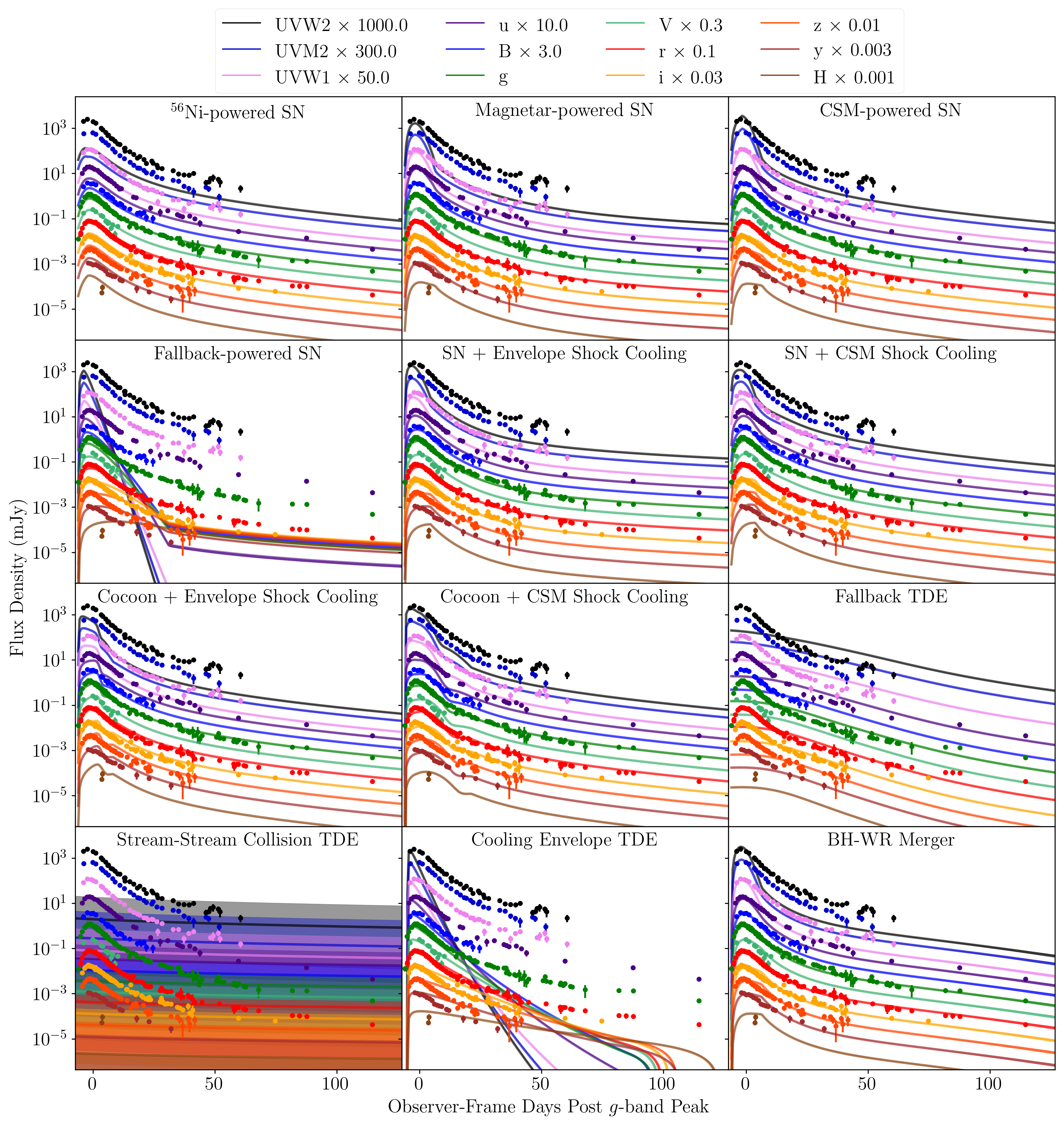}
\caption{Same as Figure \ref{fig:optfits_tfree}, but with the prior for $T_{\rm floor}$ restricted to $<$ 10$^4$ K.} %
\label{fig:optfits}
\end{figure*}

Figure \ref{fig:bbprops} shows the multiband fit from the evolving blackbody model and the luminosity, photospheric radius, and temperature derived from it; the values found by \citet{Perley2026} are also shown as a comparison.  The model provides a good fit to the overall, showing that a blackbody with power-law time dependence can approximate the system well.  The inferred luminosities, radii, and temperatures also broadly agree with other measurements from \citet{LeBaron2026}.  The radius increase with power-law index $\sigma_{\rm R, rise} = 0.01^{+0.01}_{-0.01}$ and decreases with $\sigma_{\rm R, decline} = -1.06^{+0.01}_{-0.01}$, while the temperature rises with $\sigma_{\rm T, rise} = 1.02^{+0.01}_{-0.01}$ and declines with $\sigma_{\rm T, decline} = -0.35^{+0.01}_{-0.01}$, where $R$ or $T \propto t^\sigma$.  The luminosity indices are determined from $R$ or $T$, which give an initial rise index of $4.10^{+0.06}_{-0.06}$, an intermediate decline index of $-1.38^{+0.06}_{-0.06}$, then a final decline index of $-3.52^{+0.06}_{-0.06}$.  The inferred luminosities, radii, and temperatures do differ from \citet{Perley2026} at early times, likely due to the simplified power-law time evolution of those properties in our model, so inferred parameters during the early and intermediate phases are likely unreliable.

Figure \ref{fig:optfits_tfree} shows the fits to the optical data with the more physically-motivated models, with the posteriors listed in Appendix \ref{sec:priors}.  Several models, including the magnetar-powered SN model, SN + envelope shock cooling model, and black hole-Wolf Rayet (BH-WR) merger model, can reproduce the rise and decline timescales, luminosity, and colour evolution of the light curve well.  A few other models, including the nickel-powered and fallback-powered SN models as well as the cocoon + envelope and SN + CSM shock cooling models, can somewhat reproduce the data but underpredict the UV emission around peak.  The TDE models all fail to describe the data, with the fallback and stream-stream collision models evolving much too slowly, and the cooling envelope model cooling far too quickly.

The parameters for each of the SN, shock cooling, and BH-WR merger models all contain a plateau temperature, $T_{\rm floor}$, which has a wide prior between 10$^3$ and 10$^5$ K.  Models with this parameter keep the photospheric temperature at a minimum value $T_{\rm floor}$ at late times, resulting in the photospheric radius receding as the luminosity continues to decrease.  This allows the models to approximate a time when the photosphere recedes without the need to continuously calculate the optical depth of the ejecta.  \citet{Nicholl2017} mention that this parameter may contain information about other processes keeping the ejecta at a constant temperature, such as recombination or ejecta fragmentation from a central engine.  Using $T_{\rm floor}$ also allows the model to make multiband predictions deep into the nebular phase, when the photosphere no longer exists and the emission can no longer have a blackbody spectrum.  Emission at this phase is driven mostly by line emission in supernovae, although in LFBOTs, free-free emission and non-thermal emission from the heating source may also be important.  Because of this, a blackbody approximation can not make accurate predictions past the photospheric phase \citep[see discussion in][]{Schulze2024}.

The values of $T_{\rm floor}$ from the fits shown in Figure \ref{fig:optfits_tfree} are between 20 000 -- 40 000 K, similar to the values found for the LFBOT AT2020xnd by \citet{Omand2024}.  This is much higher than the $\sim$ 5000 K that has been estimated for other transients \citep{Nicholl2017, Taddia2019}.  This implies the photosphere begins to recede early, which is consistent with the data.  However, this also implies that the ejecta becomes optically thin within $\lesssim$ 10 days (see Appendix \ref{sec:photo}), which is not consistent with the spectroscopic data \citep{Perley2026}.  

We re-fit the optical data with the same models, but with the upper limit on the prior for $T_{\rm floor}$ set to 10$^4$ K - these fits are shown in Figure \ref{fig:optfits} with the posteriors listed in Appendix \ref{sec:priors}.  $T_{\rm floor} \leq 10^4$ K was chosen because it is below the inferred late-time temperature of the LFBOT (Figure \ref{fig:bbprops}), so a model would have to self-consistently calculate the photospheric radius and temperature to explain the data.  The TDE models do not use $T_{\rm floor}$, and thus remain unchanged.  Most models now cool to $\leq$ 10$^4$ K on timescales comparable to the peak of the LFBOT or earlier, and fail to reproduce the UV data.  Only the magnetar-driven SN model and BH-WR merger model can reproduce the light curve peak, but cool down shortly after and fail to reproduce the observed decline, again underpredicting the UV data.  

\subsection{Radio and X-ray}

\begin{figure}
\hspace*{-0.5cm}
\includegraphics[width=\linewidth]{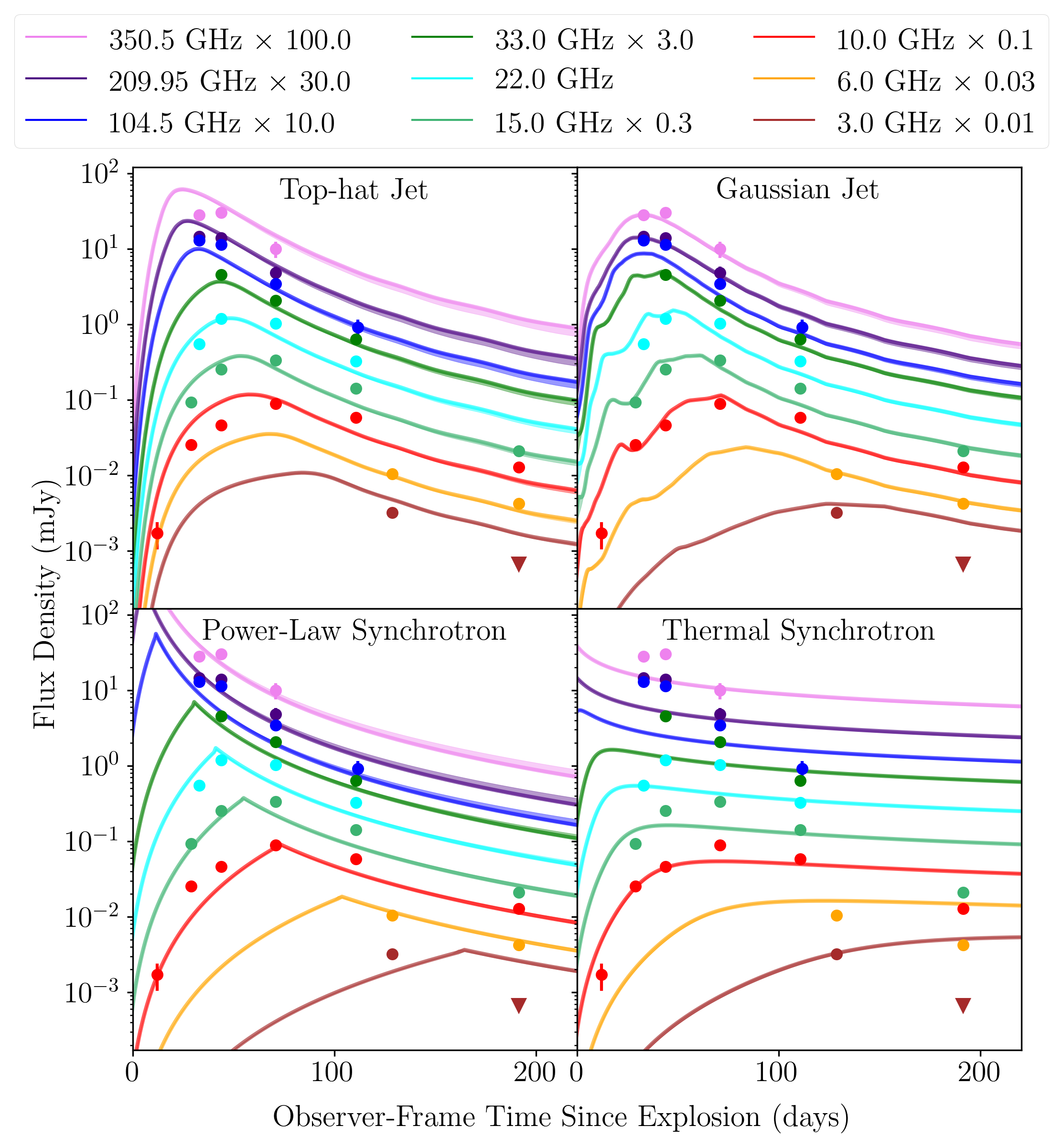} \\
\hspace*{-0.5cm}
\includegraphics[width=\linewidth]{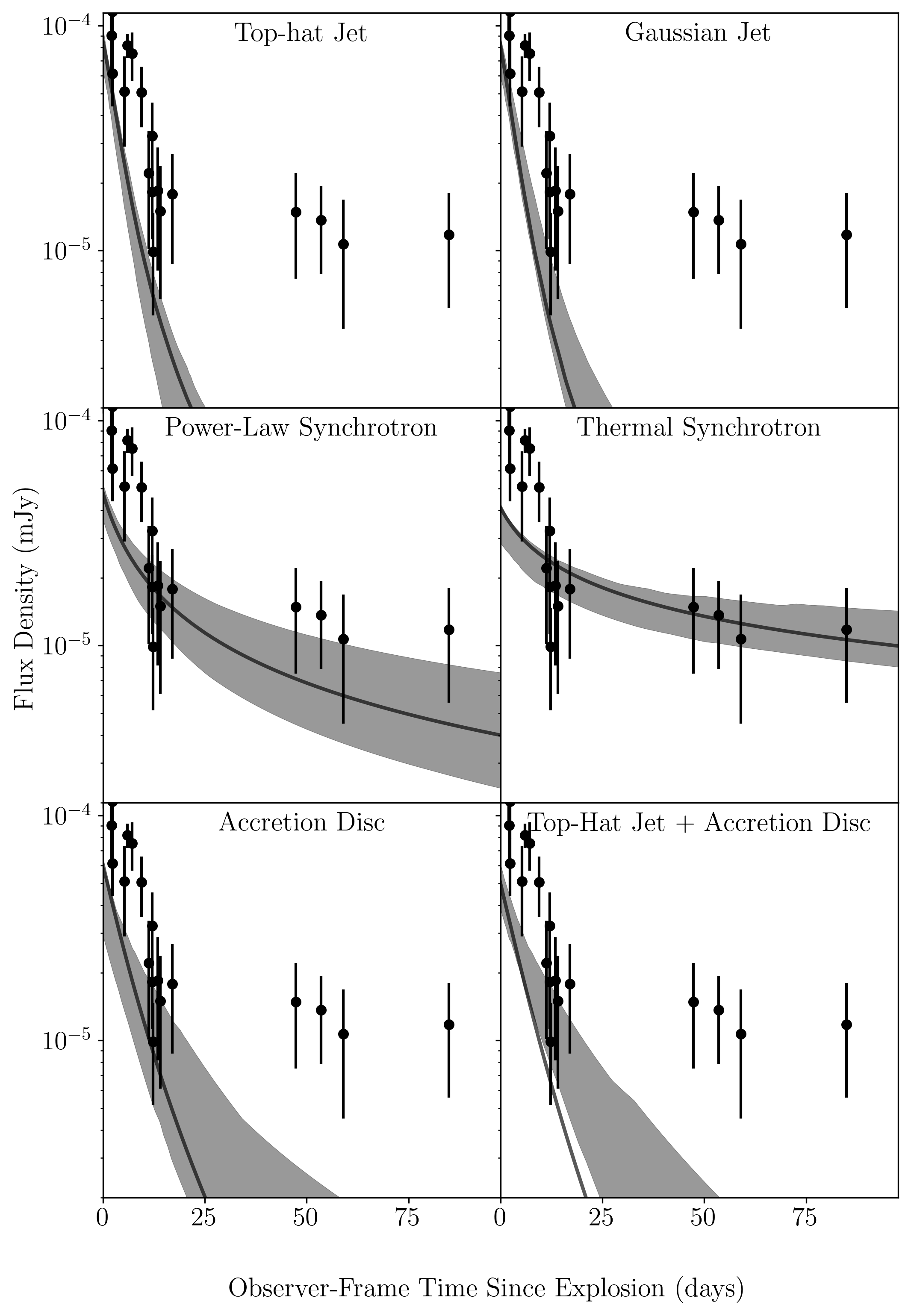}
\caption{The light curve fits to the radio and submillimetre (top) and X-ray (bottom) data for each model discussed in Section \ref{sec:mods} (see Table \ref{tbl:models}).  The solid line shows the model with the highest likelihood while the shaded region shows the 90$\%$ confidence interval.} %
\label{fig:radx_lcs}
\end{figure}

Figure \ref{fig:radx_lcs} shows the fits to the radio and X-ray data, which were modelled separately to allow for the possibility that they could arise from separate processes.  The posteriors are listed in Appendix \ref{sec:priors}.  The radio data are better reproduced by the two afterglow models and the power-law synchrotron blast wave model compared to the thermal synchrotron blast wave model.  Specifically, the afterglows and power-law synchrotron show a more pronounced peak on the same timescale as the data and a faster decline, although none of the models can reproduce the fast decline consistent with the 3 GHz upper limit.  The jets are both predicted to be strongly off-axis, with $\theta_{\rm obs}/\theta_{\rm core} \sim 5-10$.  The power-law synchrotron is predicted to be in a medium with $\rho \approxprop r^{-2.9}$.  The fits to the X-ray data show a different preference; the synchrotron blast wave models reproduce the later data well but underpredict the early data, while the jet and accretion disc models fail to reproduce the data entirely, especially at late times.

\begin{figure}
\includegraphics[width=\linewidth]{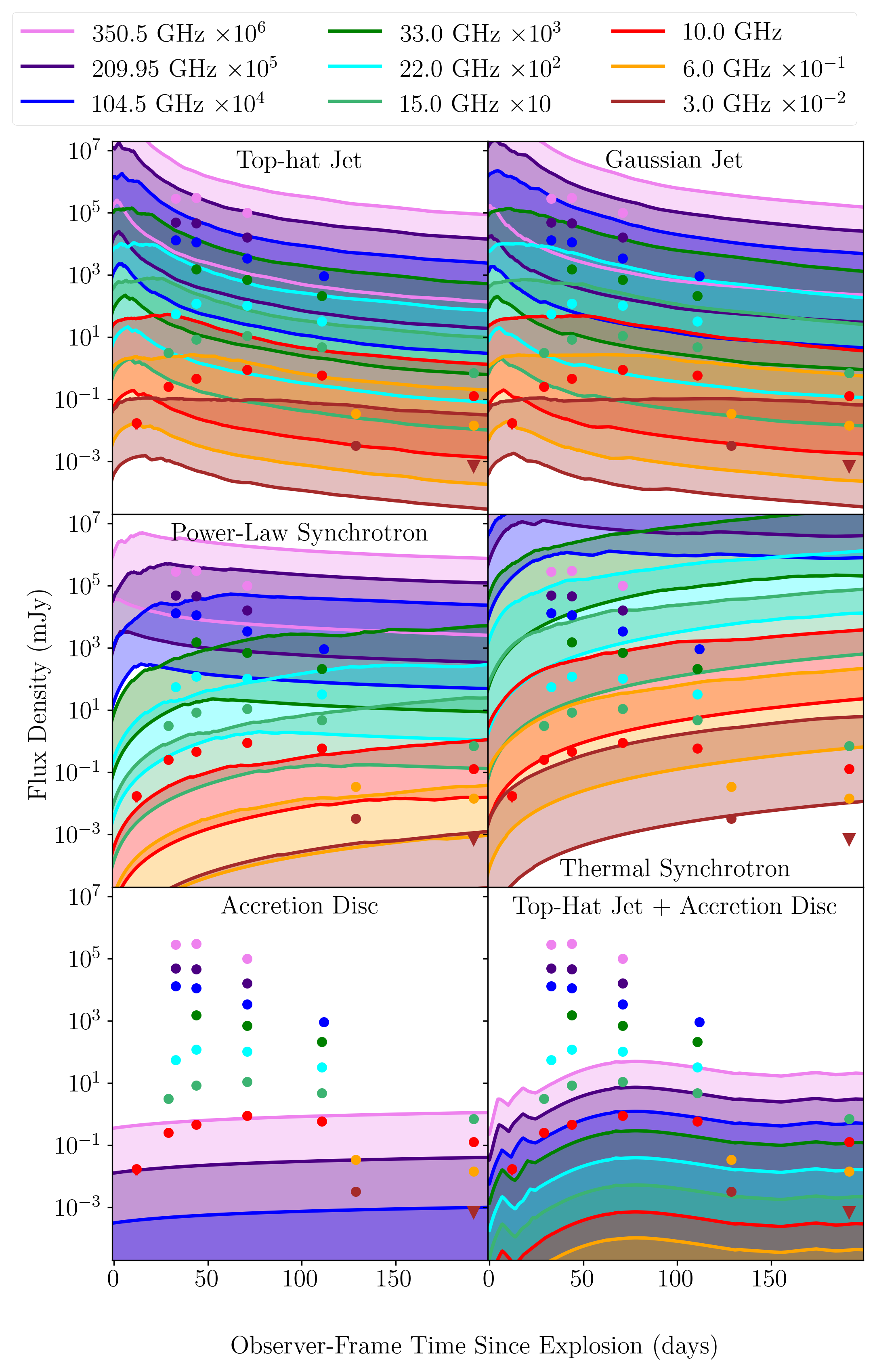} \\
\includegraphics[width=\linewidth]{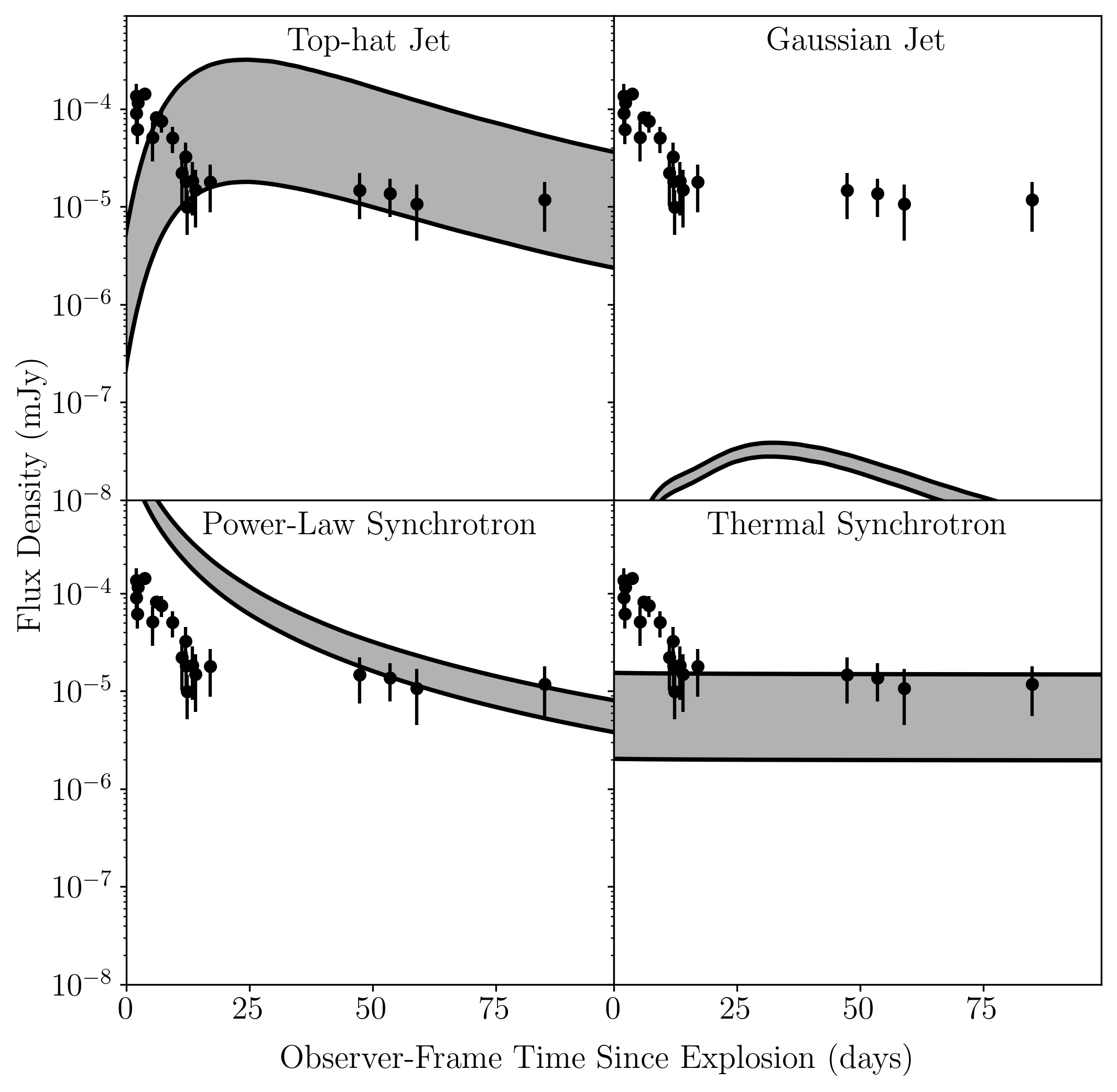}
\caption{Radio light curves using the posteriors from the X-ray fits (top) and X-ray light curves using the radio posteriors (bottom) (see Figure \ref{fig:radx_lcs} for the fits). The shaded region shows the 90$\%$ confidence intervals, while the solid lines show the upper and lower bounds of the 90$\%$ confidence intervals.} %
\label{fig:radx_posts}
\end{figure}

An important consideration from the radio and X-ray fits is what they predict for the opposite band, since a model that overpredicts the other band would not be a viable candidate.  Figure \ref{fig:radx_posts} shows the fits presented in Figure \ref{fig:radx_lcs} extended into the opposite band.  The two jet models and power-law synchrotron model from the X-ray data all have wide confidence intervals that roughly overlap with the radio data, while the thermal synchtrotron model overpredicts the radio data and the accretion disc and jet + accretion disc models underpredict.  The models that are consistent with the radio data show very different behaviours in the X-ray, with the Gaussian jet severely underpredicting the data; thermal synchrotron being consistent with the late data but underpredicting the early data; the top-hat jet being roughly consistent but showing a rise on the timescale of 25 days, which is not observed; and the power-law synchrotron model overpredicting the data by an order of magnitude.

Putting these results together yields no obvious explanation for the non-thermal emission.  The radio could be explained by an off-axis Gaussian jet, since the X-ray flux of that model does not overpredict the data.  The X-ray flux is only consistent with the synchrotron models, but the thermal synchrotron overpredicts the radio data.  The power-law synchrotron models from the X-ray fits are roughly consistent with the radio data, yet the models from the radio fits overpredict the X-ray data by more than an order of magnitude.  If the X-rays are powered by a power-law synchrotron blast wave, there may be emission from a second component, like an off-axis jet, that can better reproduce the early radio emission, but that component would need to have a radio flux not significantly larger than the blast wave and a negligible X-ray flux.

\begin{figure}
\hspace*{-0.5cm}
\includegraphics[width=1.1\linewidth]{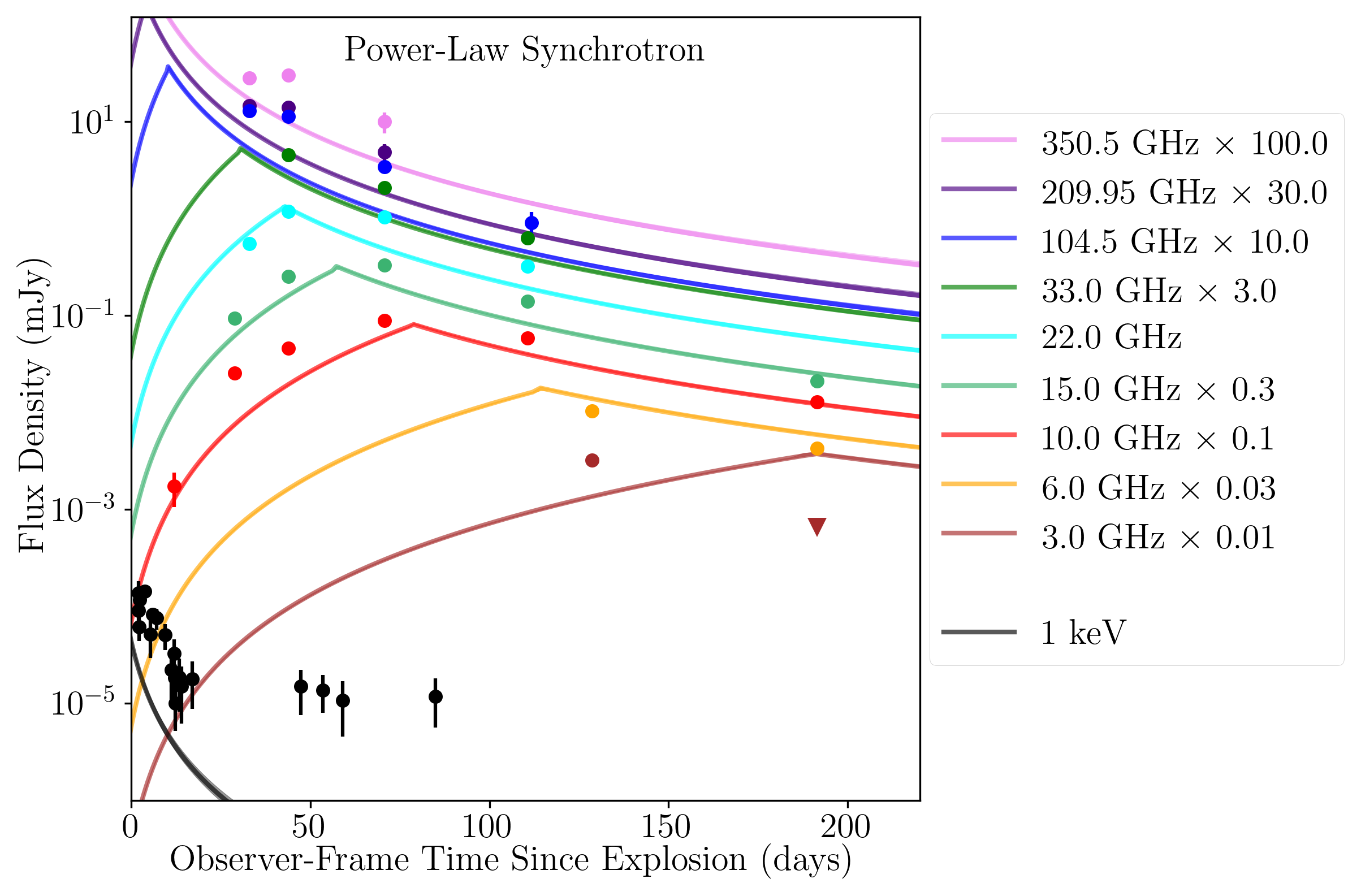} \\
\hspace*{-0.5cm}
\includegraphics[width=1.1\linewidth]{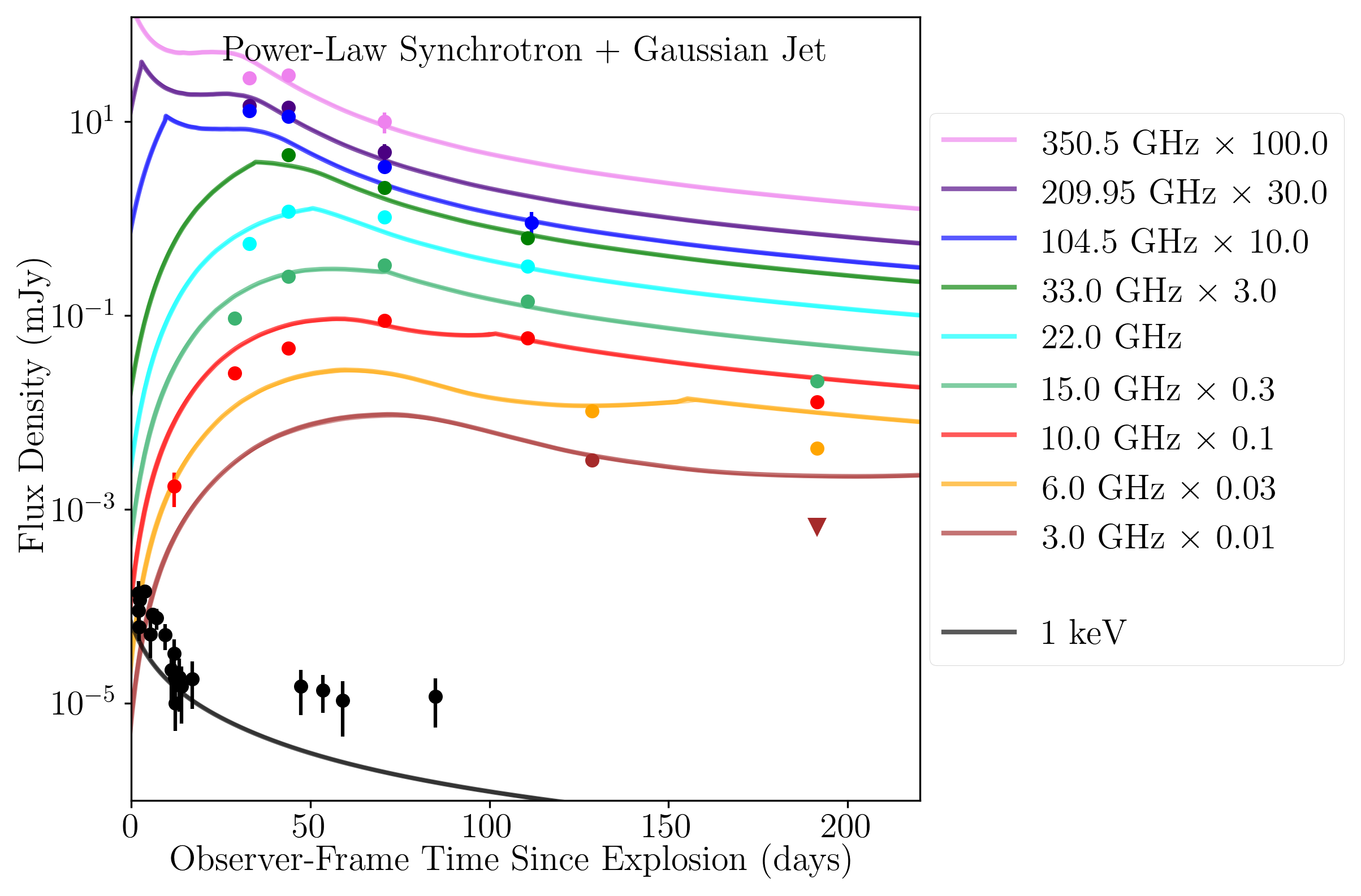}
\caption{The light curve fits to the radio, submillimetre, and X-ray data for the power-law synchrotron (top) and thermal synchrotron (bottom) models.  The solid line shows the model with the highest likelihood while the shaded region shows the 90$\%$ confidence interval.}
\label{fig:radx_combined_lcs}
\end{figure}

The two possibilities to explain the non-thermal emission that arise from our analysis, assuming that the data are not described by a model we have not investigated, are that a power-law synchrotron blast wave can explain both the radio and X-ray data, or that a combined Gaussian jet and power-law synchrotron blast wave are needed to explain the data.  We show the combined radio and X-ray fits for both of these scenarios in Figure \ref{fig:radx_combined_lcs}, with the posteriors listed in Appendix \ref{sec:priors}.  The power-law synchrotron fit is similar to the fits presented in Figure \ref{fig:radx_lcs}: The model underpredicts X-rays at early times but reproduces the data at late times, and does not reproduce the peak or the decline in the radio data.  The synchrotron + Gaussian jet model reproduces the peak and decline in the radio data, but underpredicts the X-ray data.  While neither of these possibilities can explain the data, this might be due to these models assuming a power-law density profile for the CSM.  A more complicated profile may be able to explain the differences at late time for the radio (see discussion in Section \ref{sec:otherpapers}), although it seems likely that the X-ray emission was produced by a process we have not considered here.

\section{Discussion} \label{sec:disc}

\subsection{Model Analysis and Late-Time Predictions}

The optical models tested all failed to reproduce the optical emission unless the temperature was allowed to plateau at $>$ 20 000 K.  This highlights the need for multiband fitting when examining light curve models, as opposed to bolometric fitting.  After the diffusion phase, a constant temperature implies $L \propto R_{\rm phot}^2$, which may be a good approximation when the photosphere is receding. The recession of the photosphere in an expanding medium begins when the optical depth of the ejecta drops fast enough to counteract the effect of the ejecta expansion.  This timescale can be connected to the nebular timescale, when the entire ejecta becomes optically thin.  Appendix \ref{sec:photo} shows that the ratio between these two timescales for a homologously expanding medium with any non-increasing density profile is $< \sqrt{3}$.  While the data show the photospheric radius peaks on a timescale of $\sim$ 5 days, spectra at $\sim$ 40 days still show the emission being dominated by a blackbody-like continuum \citep{Perley2026}.  Since this is inconsistent with the derived timescales, this suggests that the ejecta responsible for the thermal emission cannot be homologously expanding, which rules out the SN, shock cooling, and BH-WR merger models.  This can also rule out some models we did not test, such as the TDE-powered SN model from \citet{Tsuna2025} or the Helium core-black hole merger from \citet{Klencki2025}.  This does not rule out the presence of homologous components that may produce spectral lines or non-thermal emission, but these components cannot be the primary source of the optical emission.  This photospheric behaviour also does not rule out TDEs, although none of the TDE models were able to explain the optical emission.  However, TDE models are known to be uncertain and the emission processes of supermassive black hole (SMBH) TDEs, especially around peak are not well understood \citep[e.g.][]{Sarin2024CE, Wise2026, Angus2026}, so the possibility that this transient could be an IMBH TDE cannot be completely discounted.  The lack of early rest-frame optical/near-UV spectral features may point to AT2024wpp as a low mass analogue to featureless TDEs \citep{Hammerstein2023}. However, the presence of the dense extended material inferred from radio emission is not predicted by any current TDE models.

\begin{figure}
    \centering
    \includegraphics[width=\linewidth]{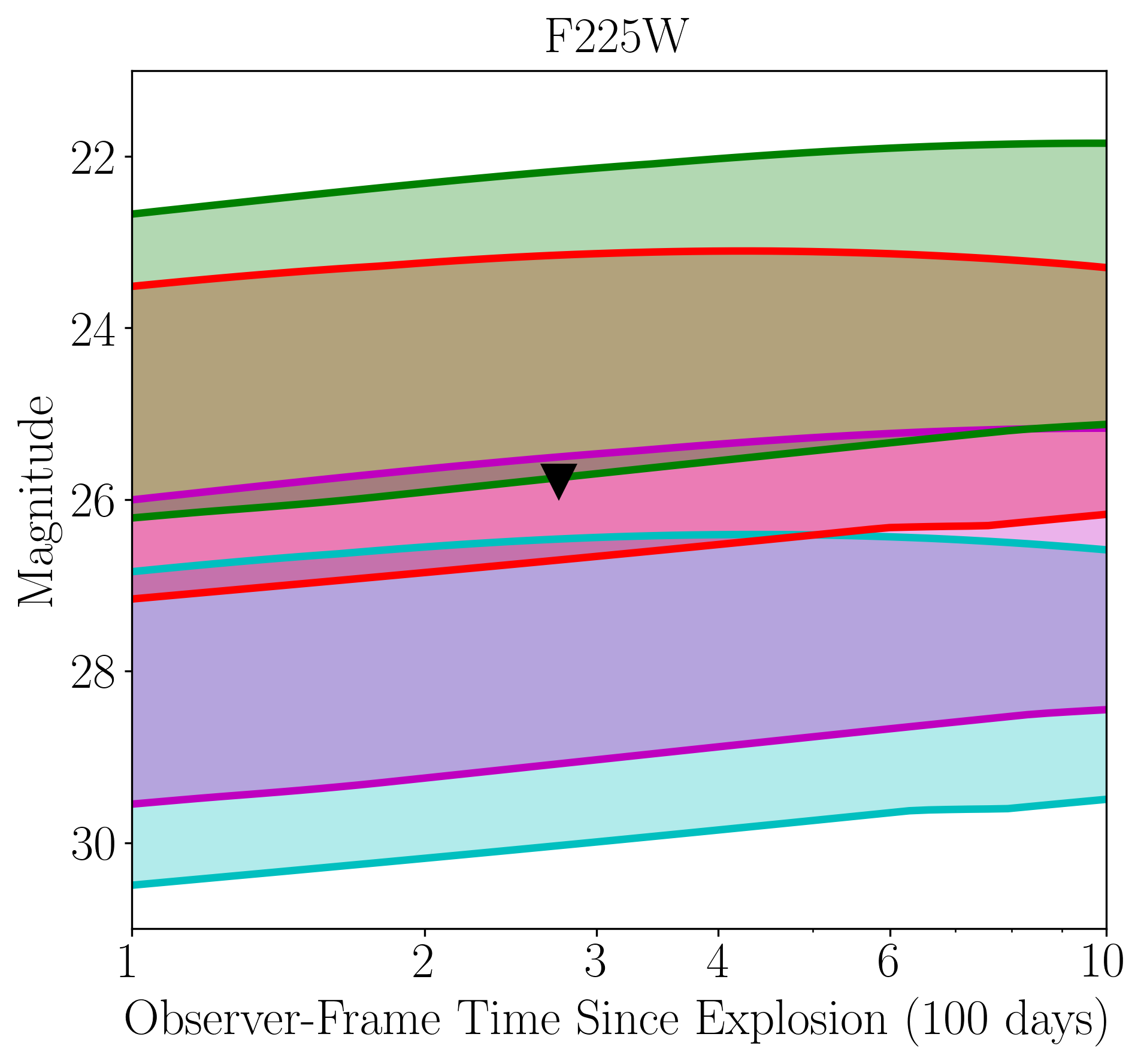} \\
    \includegraphics[width=\linewidth]{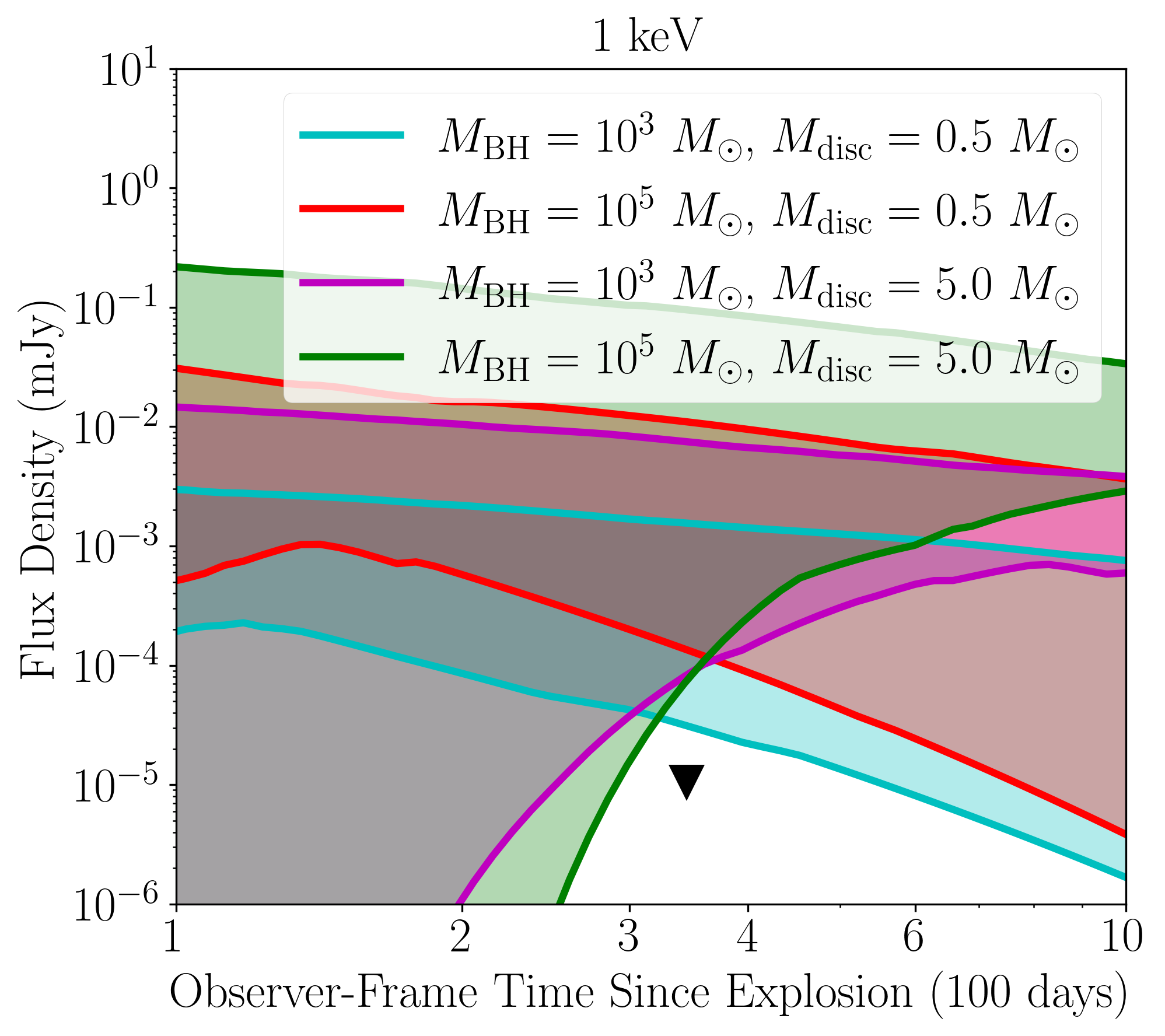}
    \caption{The predicted late-time emission from the accretion disc of an IMBH TDE.  Each colour represents a different black hole mass and disc mass, while the width of the band shows the 90$\%$ confidence interval from marginalizing over the default priors for the other parameters (see \citet{Mummery2025} for details).  The solid lines show the upper and lower bounds of the 90$\%$ confidence interval.  The top panel shows the prediction for F225W, while the bottom panel shows the predictions at 1 keV. The triangles denote the observed 3$\sigma$ upper limits.}
    \label{fig:tde_plateaus}
\end{figure}

One way of testing models is using their early properties to predict emission at later times \citep[e.g. ][]{Law2019, Omand2023, Margutti2023}.  The late-time emission from tidal disruption events can be predicted in radio \citep[e.g. ][]{Zauderer2011, vanVelzen2016, Andreoni2022, Alexander2026}, infrared \citep[e.g. ][]{Dou2016, Newsome2024, Masterson2024}, X-rays \citep[e.g. ][]{Bade1996, Donley2002, Grotova2025}, and may also be in neutrinos \citep[e.g. ][]{Stein2021, Jiang2023, Ji2025, Toshikage2025}.  One prediction from all TDE models is that the system will settle into an accretion disc at late times, which will emit primarily in optical/UV and X-rays \citep{Mummery2025}.  Since the TDE models for early optical emission did not reproduce the data for AT2024wpp, we do not use the posteriors from those models for predictions.  We instead present F225W UV and 1 keV predictions\footnote{These predictions were made using the \textsc{GR\_disc} model from \textsc{FitTeD} \citep{Mummery2025}, which ignores relativistic photon deflection and gravitational redshift.  Marginalizing over a prior using a model incorporating these effects (\textsc{GR\_disc\_plus}) would be computationally unfeasible.  These effects are not expected to strongly affect optical/UV emission, but may introduce uncertainty in the X-ray predictions.} for 10$^3$ and 10$^5$ M$_\odot$ black holes and 0.5 and 5.0 M$_\odot$ accretion discs in Figure \ref{fig:tde_plateaus}.  These black hole masses were chosen because the plateau emission from AT2018cow was estimated to originate from a $\sim$ 10$^3$ M$_\odot$ black hole \citep{Inkenhaag2023, Inkenhaag2025}, with 10$^5$ M$_\odot$ acting as a conservative upper limit. The observer-inclination angle and black hole spin are marginalized over using their default priors \citep{Mummery2025}, while the initial disc radius $r_0$ and viscous timescale $t_{\rm visc}$ are calculated via

\begin{equation}
    r_0 = 2R_\star(M_{\rm BH}/2M_{\rm disc})^{1/3}
    \label{eqn:r0}
\end{equation}
and 

\begin{equation}
    t_{\rm visc} = \mathcal{V} \sqrt{r_0^3/GM_{\rm BH}} .
    \label{eqn:tvisc}
\end{equation}
We use the mass-radius relation $R_\star = (M_\star/M_\odot)^{4/5}R_\odot$ appropriate for lower-mass main sequence stars \citep{Metzger2022CE, Sarin2024CE}, and the parameter $\mathcal{V}$ encapsulates the poorly understood MHD effects in the disc physics.  We sample $\mathcal{V}$ with a log-uniform prior between $10^2$ and $10^4$. Shown also are  3$\sigma$ upper limits from \citet{Perley2026}.

The F225W predictions have $\sim$ 3 mag of uncertainty, with the 10$^5$ M$_\odot$ black holes having emission between 22-26th mag and the 10$^3$ M$_\odot$ black holes having emission between 26-30th mag, making it likely this emission could be detected if the black hole mass is close to the IMBH-SMBH boundary, but more parameter dependent if the black hole mass is lower.  The X-ray predictions have $\gtrsim$ an order of magnitude of uncertainty, especially during the first year, but are predicted to be between $\sim$ 10$^{-1}-10^{-5}$ mJy over the next two years.  More massive black holes and discs have higher early uncertainty but higher luminosity at late times.  The HST UV upper limit disfavours black holes with masses $\gtrsim 10^5$ $M_\odot$, but does not strongly constrain discs around lower mass black holes.  The \textit{Swift} X-ray upper limit at $\sim$ 1 year strongly disfavours these all of these plateau models, which suggests either a lower mass disc or black hole, or that AT2024wpp is not an IMBH TDE.  

\begin{figure}
    \centering
    \includegraphics[width=\linewidth]{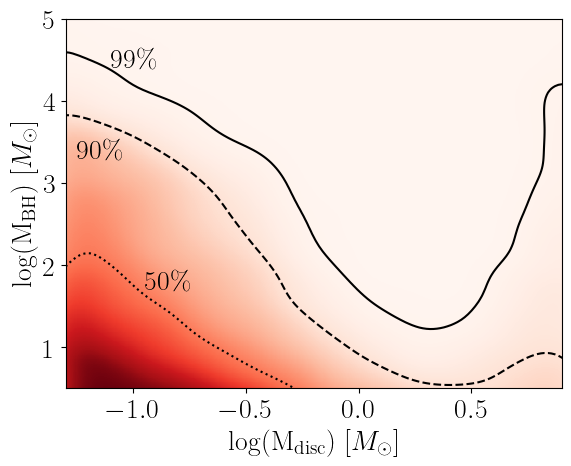} \\
    \caption{The viable region for plateau emission given the late-time UV and X-ray upper limits in an IMBH TDE.  The darker regions indicate a high posterior density.  The dotted, dashed, and solid lines indicate parameter regions excluded at 50\%, 90\%, and 99\% confidence, respectively.}
    \label{fig:xraylim_tde}
\end{figure}

The late-time optical, UV, and X-ray upper limits can be used to exclude some regions of the $M_{\rm BH} - M_{\rm disc}$ parameter space through modelling of the plateau emission in the case of an IMBH TDE - this is shown in Figure \ref{fig:xraylim_tde} for all upper limits at $>$ 200 days, including limits from the VLT, HST, and \textit{Swift}.  We sample a log-uniform prior for both $M_{\rm BH}$ and $M_{\rm disc}$, with $M_{\rm BH}$ ranging from $1 - 10^5 $ $M_\odot$ and $M_{\rm disc}$ ranging from $0.05 - 10$ $M_\odot$.  The upper limit for the $M_{\rm BH}$ prior is because any larger black hole would cause the initial transient to be too slow, and the lower limit on disc mass prior was chosen so we could examine tidal disruptions of the lowest mass stars. We calculate $r_0$ and $t_{\rm visc}$ using Equations \ref{eqn:r0} and \ref{eqn:tvisc}, and all other parameters use their default priors.  Disc masses $\gtrsim$ 1 $M_\odot$ are excluded with 99\% confidence for a black hole mass of $> 10^2$ $M_\odot$.  At 90\% confidence, disc masses $\gtrsim$ 0.3 $M_\odot$ are excluded for a black hole mass of $> 10^2$ $M_\odot$.  Black hole masses of $\gtrsim 10^{4.5}$ and $\gtrsim 10^{4}$ $M_\odot$ are excluded for all disc masses at 99\% and 90\% confidence respectively.  However, we caution against drawing too strong a conclusion from a posterior based only on a few upper limits.

\subsection{Other Scenarios}

\citet{Metzger2022BHWR} outlined several progenitor scenarios for LFBOTs, including engine-powered SNe \citep[e.g.][]{Prentice2018}, pulsational pair instability SNe \citep[PPISNe, e.g.][]{Leung2021}, prompt collapse from failed SNe or PPISNe \citep[e.g.][]{Margutti2019}, IMBH TDEs \citep[e.g.][]{Quataert2019, Perley2019, Margutti2019}, stellar mass BH TDEs \citep[e.g.][]{Perets2016, Kremer2019, Kremer2023, Beniamini2025}, failed common envelope with a prompt merger \citep[e.g.][]{Soker2019a, Soker2019b, Schroder2020}, and failed common envelope with a delayed merger \citep[e.g.][]{Metzger2022BHWR}.  This work examined engine-driven SNe and TDEs, while PPISNe and current failed common envelope models are not supported by the photospheric radius timescales derived in Appendix \ref{sec:photo}.  

Failed SNe remain an intriguing possibility, though predicting their optical signatures is extremely challenging \citep{Beasor2024}, as theoretical models offer a wide range of predictions for what a failed SN might look like.  Different models of failed SNe have predicted no observable signature \citep{Kochanek2008}, a long-lived infrared transient with little to no optical emission \citep{Lovegrove2013}, a long-lived optical transient \citep{Perna2014}, or even a short-lived relatively bright transient \citep{Antoni2023, Antoni2025}, similar to an LFBOT.
\citet{Chrimes2026} examine the possibility of failed SNe from very massive stars ($>$30 $M_\odot$), similar to the progenitors of super-kilonovae \citep{Siegel2022}, and find that the rates, host metallicities, mass loss rates, and expected plateau emission roughly match with the LFBOT population and previous observations.  In this case, LFBOTs may contribute significantly to r-process enrichment in galaxies.

\begin{figure}
    \centering
    \includegraphics[width=\linewidth]{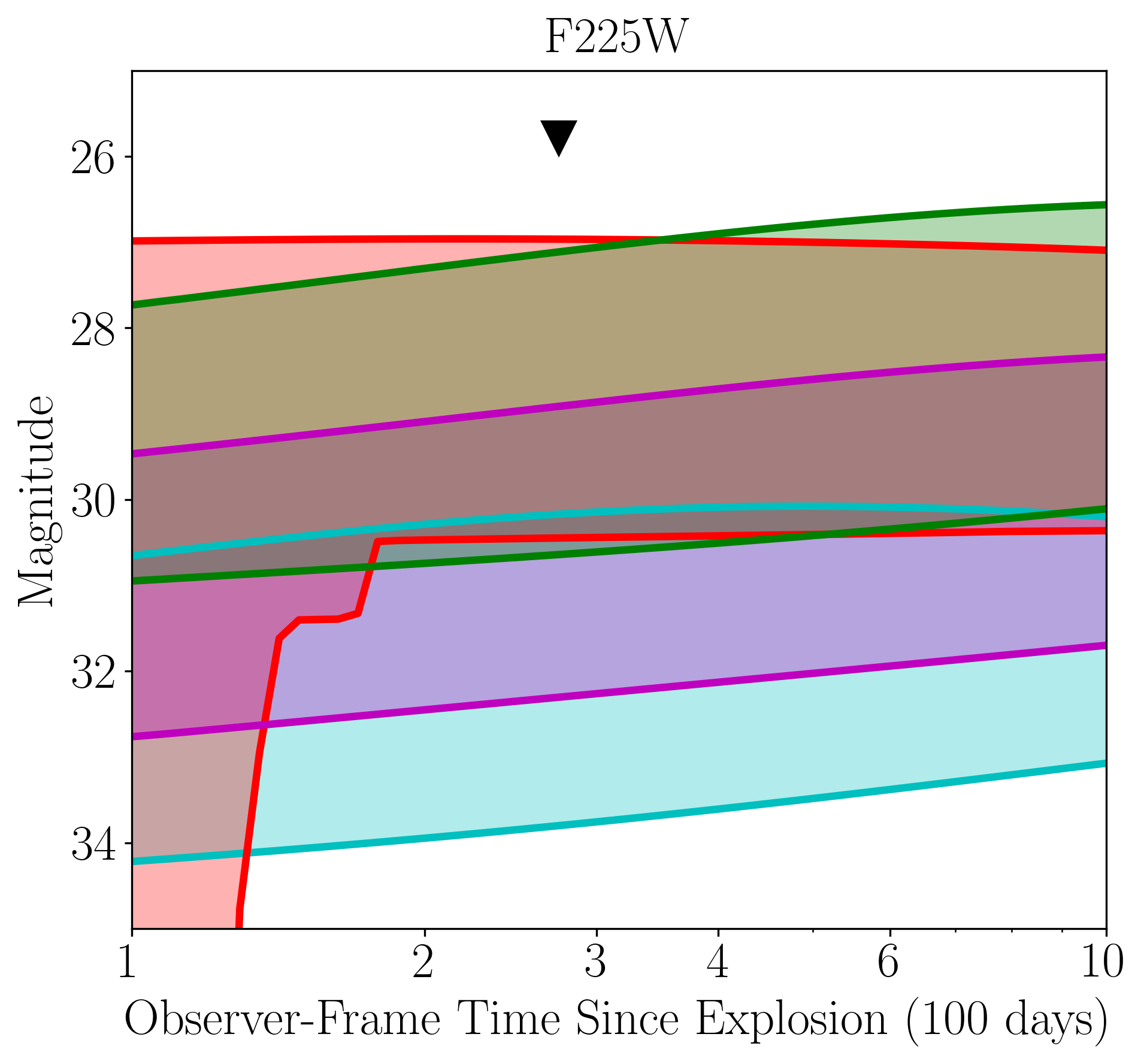} \\
    \includegraphics[width=\linewidth]{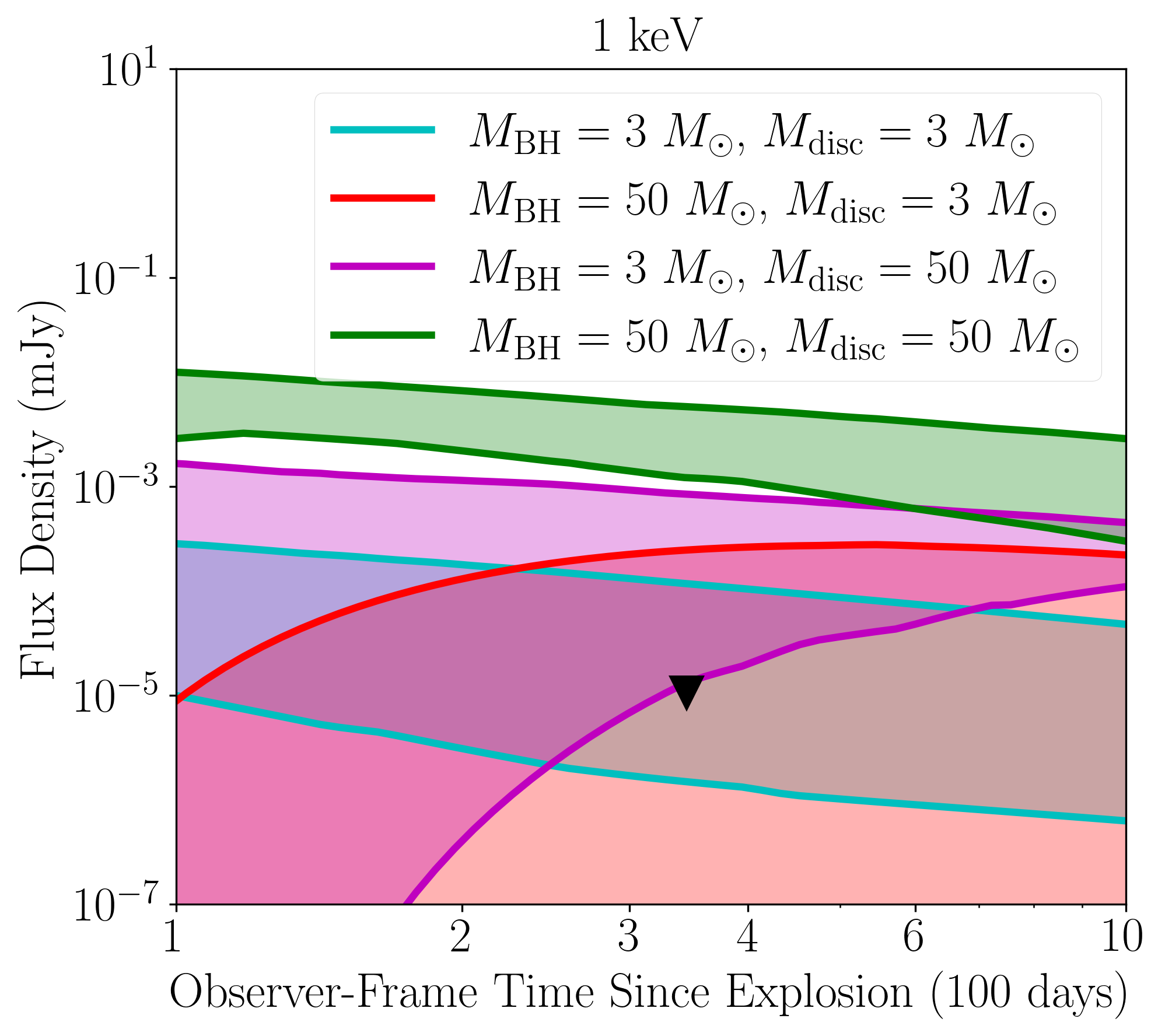}
    \caption{Same as Figure \ref{fig:tde_plateaus}, but for discs resulting from failed SNe.}
    \label{fig:fsn_plateaus}
\end{figure}

Although there are no reliable models for the optical flare from a prompt collapse failed SN, the system should settle into an accretion disc at later times, similar to TDEs.  The expected plateau emission for failed SNe involving high and low mass stars are shown in Figure \ref{fig:fsn_plateaus}.  The priors are similar to the IMBH TDE plateaus, except the initial disc radius $r_0$ is calculated by assuming the progenitor star is maximally rotating and the angular momentum of the black hole $J_{\rm BH}$ is much less than the angular momentum of the disc $J_{\rm disc}$.  These assumptions give a radius

\begin{equation}
    r_0 = 2 R_\star \left({M_\star \over M_{\rm disc}}\right)^2\left({M_\star \over M_{\rm BH}}\right) ,
\end{equation}
with $M_\star = M_{\rm disc} + M_{\rm BH}$.  It is worth noting that this calculation ignores the self-gravity of the disc, which can be important when $M_{\rm disc} \gtrsim M_{\rm BH}$.  The F225W emission has $\sim$ 3 mag of uncertainty and is fainter than the IMBH TDE plateau, and the observed upper limit is not constraining.  The X-ray emission is predicted to be fainter than 10$^{-2}$ mJy but have lots of uncertainty, except for the $M_{\rm disc} = M_{\rm BH} = 50 M_\odot$ case.  The X-ray upper limit disfavours discs $\gtrsim$ 50 $M_\odot$ but does not constrain black hole mass.  Sampling the prior to find the excluded area in the failed SN scenario, as done with TDEs in Figure \ref{fig:xraylim_tde}, does not give meaningful constraints on either the black hole or disc mass.

Another scenario of interest is a reprocessing outflow, where an accretion disc powers an outflow that reprocesses the disc emission, powering an optical transient \citep{Piro2020, Uno2020fbot}.  Models of these outflows have successfully reproduced previous TDEs \citep{Uno2020tde, Matsumoto2021} and LFBOTs \citep{Uno2020fbot, Uno2023}.  The continuous mass injection of the model means that even though the outflow is expanding, the properties of the photospheric radius cannot be described by the relations in Appendix \ref{sec:photo}.  The photospheric radius will expand outward initially with the outflow until it reaches the radius where the outer region of the outflow is too diffuse to absorb light, and its recession is governed by the time-dependence of the mass outflow rate, with the outflow becoming optically thin only when the outflow rate becomes sufficiently small.  This should lead to an early photospheric peak and a late nebular phase, which is observed in AT2024wpp. Both \citet{Piro2020} and \citet{Uno2023} predict an increase in photospheric temperature within the first 10 days post-explosion when the mass outflow rate and central luminosity are coupled, contrary to what is observed, but \citet{Piro2020} show that the optical emission from AT2018cow can be reproduced if the central luminosity decreases faster than the mass outflow rate; they speculate that this could arise from either the X-rays being produced by a wind interacting with an equatorial torus which erodes over time, or from a viscously spreading disk causing the outflow velocity to decrease over time.  A calculation of the X-ray opacity within this model would be useful to predict when the emission from the accretion disc will escape unattenuated.

\subsection{Summary of Previous Interpretations} \label{sec:otherpapers}

Our analysis of AT2024wpp and our interpretation of the event is mostly in broad agreement with previous studies.  \citet{LeBaron2026} and \citet{Perley2026} both concluded that a homologously expanding medium can not explain the optical transient, and preferred a reprocessing outflow model similar to that discussed above, although no attempts were made to compare a model light curve to the data. \citet{Aspegren2026} reach a similar conclusion through simulations of line formation and suppression in LFBOT- and TDE-like transients.  \citet{LeBaron2026} also discuss a dense CSM interaction model \citep{Khatami2024} as well as BH-WR merger and TDE-powered SN models \citep{Metzger2022BHWR, Tsuna2025}, ruling out the former but not the latter two.  

The density profile and density from our modelling of the radio emission with a power-law synchrotron blast wave were $n_{\rm CSM} \approxprop r^{-2.9}$ and $n_{\rm CSM} \approx 10^7$ cm$^{-3}$ at 10$^{16}$ cm respectively, which match well with the shock modelling from \citet{Nayana2025} and \citet{Perley2026}.  Those works find that the shock radius recedes after $\sim$ 100 days, something which is not accounted for in any of our models and can explain the discrepancy at late times.  They also both point out that these CSM properties seem to be typical of LFBOTs as well as some interacting SNe, such as the Type Ibn SN 2006jc \citep{Maeda2022}.  This may imply similar progenitors for the two systems, as previously suggested by \citet{Metzger2022BHWR} and \citet{Klencki2025}.  

\citet{LeBaron2026} observe an excess of IR emission after $\sim$ 20$-$30 days, which was not well described by \citet{Perley2026} due to a lack of late-time IR observations.  This excess is similar to the one observed in AT2018cow \citep{Perley2019, Margutti2019}, although it emerges at a later timescale.  Two models that can explain the IR excess are free-free opacity effects, which require several solar masses of outflow \citep{Chen2022}, and a dust echo \citep{Metzger2023}, which reproduces the emission with densities similar to those inferred from the radio emission.  As noted by \citet{LeBaron2026}, this means the radio and IR emission can form in the same medium by completely separate processes.

\citet{Nayana2025} found that the X-ray rebrightening and spectral steepening can be explained by a Compton bump, and posit that the timescale may be due to a drop in the X-ray opacity due to ionization breakout \citep{Metzger2014, Metzger2014kn} at $\sim$ 40 days.  This also roughly corresponds to the timescale at which the broad features disappear from the optical spectrum \citep{Perley2026}.  The appearing and disappearing features are expected in a reprocessing outflow model, as the early medium should be hot and ionized until enough material has been expelled, then should remain partially ionized at the photosphere until the photosphere recedes into the ionized region (analogous to a Str\"{o}mgren sphere) caused by the inner X-ray source.  \citet{Nayana2025} favour a model with super-Eddington accretion onto a compact object which produces aspherical disc-wind outflows.  The polarization measurements by \citet{Pursiainen2025} seem to prohibit large scale asymmetries in the outflow, although small-scale clumping can produce a polarization signal below the observed noise threshold \citep[e.g.][]{Tanaka2017}.

Finally, \citet{Perley2026} used scaling relations \citep{Mummery2024scaling} to get a black hole mass constraint from the late-time non-detections of a plateau.  They find a black hole mass limit of $3 \times 10^4 M_\odot$, similar to what we infer at 99\% confidence from Figure \ref{fig:xraylim_tde}.

\section{Summary and Conclusions} \label{sec:conc}

In this work, we examined the optical, radio, and X-ray emission from the LFBOT AT2024wpp.  The data were compared to a number of models to try and determine which physical scenario can best explain the transient.  A few optical models, including the magnetar-powered SN model, SN + envelope and SN + CSM shock cooling models, and BH-WR merger model, can reproduce the optical data if the temperature of the model plateaus above 20 000 K.  This is not unphysical but does imply that the ejecta should become optically thin on a much faster timescale than what is suggested by the spectroscopic data~\citep{Perley2026}.  While GRB afterglow models and synchrotron blast wave models can explain the radio and X-ray emission separately, a combination of these models fails to adequately explain the non-thermal emission from the system.

Analysis of the behaviour of the photospheric radius disfavours models featuring a homologously expanding ejecta.  Although the TDE models were not able to reproduce the data, we do not rule out this scenario due to the uncertainty of the emission mechanism in the stellar-mass/IMBH regime.  The late-time optical emission from an accretion disc would likely be undetectable unless the black hole mass is almost supermassive, but X-rays may be detectable for the first several years.  If the system is a TDE, current constraints favour the disruption of a low-mass star from a black hole with either intermediate or stellar mass.

AT2024wpp could be a failed SN with prompt black hole formation, but there are no reliable models for the optical flare in order to test the scenario.  A failed SN would have an accretion disc that would also be detectable in X-rays and display different behaviour than accretion discs from IMBH TDEs, allowing the scenarios to be differentiated.  A reprocessing wind may be able to explain the optical emission, and opacity calculations from that model would be useful for predicting late X-ray emission from the disc.

\section*{Acknowledgements}

The authors would like to thank Kim Page for her help with \textit{Swift} data, Ashley Chrimes and Christopher Irwin for helpful discussions, and the anonymous referee for their helpful comments.  CMBO, GPL, and CT acknowledge support from the Royal Society (grant Nos. DHF-R1-221175 and DHF-ERE-221005).  NS acknowledges support from the Kavli Foundation. HH acknowledges support from the JSPS Grant-in-Aid for Scientific Research (Grant Nos. 26H00817 and 26K07147).  ERB is supported by a Royal Society Dorothy Hodgkin Fellowship (grant no. DHF-R1-241114).  NMK acknowledges support from LIV.INNO (STFC grant ST/W006766/1).

\section*{Data Availability}

All data used are publicly available from \citet{Perley2026}.  All models used are publicly available within \textsc{Redback} \citep{Sarin_redback}.



\bibliographystyle{mnras}
\bibliography{ref} 





\appendix

\section{Priors and Summary Statistics} \label{sec:priors}

Tables \ref{tbl:sn_ni}-\ref{tbl:synchgauss_joint} show the priors and posterior summary statistics for the fits shown in Figures \ref{fig:optfits_tfree}, \ref{fig:optfits}, \ref{fig:radx_lcs}, and \ref{fig:radx_combined_lcs}.

\begin{table*}
\begin{tabular}{|cccc|cc|}
\hline
Parameter & Symbol & Units & Prior & \multicolumn{2}{c|}{Posterior} \\
& & & & Figure \ref{fig:optfits_tfree} & Figure \ref{fig:optfits} \\
\hline
Ejecta mass & $M_{\rm ej}$ & $M_\odot$ & L[10$^{-4}$, 10$^2$] & 8.83 $\pm$ 0.07 & 4.87 $\pm$ 0.00 \\
Ejecta velocity & $v_{\rm ej}$ & km s$^{-1}$ & L[10$^3$, 10$^5$] & L($5 \pm 0.00$) & L(${4.99} \pm 0.00$) \\
Nickel fraction & $f_{\rm Ni}$ &  & L[10$^{-3}$, 1] & 1.00 $\pm$ 0.00 & 1.00 $\pm$ 0.00 \\
Opacity & $\kappa$ & cm$^2$ g$^{-1}$ & U[0.05, 2] & 0.05 $\pm$ 0.00 & 0.05 $\pm$ 0.00 \\
Non-thermal opacity & $\kappa_\gamma$ & cm$^2$ g$^{-1}$ &  L[10$^{-4}$, 10$^4$] & 0.01 $\pm$ 0.00 & 0.02 $\pm$ 0.00 \\
Plateau temperature & $T_{\rm floor}$ & K &  L[10$^3$, 10$^5$] (Figure \ref{fig:optfits_tfree}) & 15110 $\pm$ 40 & \\
& & & L[10$^3$, 10$^4$] (Figure \ref{fig:optfits}) && 9998.9 $\pm$ 0.1 \\
\hline
\end{tabular}
\caption{Priors and summary statistics for the $^{56}$Ni-powered supernova supernova models in Figures \ref{fig:optfits_tfree} and \ref{fig:optfits}. Priors are either uniform (U) or log-uniform (L). The values shown for the
posterior are the median and 1$\sigma$ uncertainties. Posterior values denoted with L are given in log-space.}
\label{tbl:sn_ni}
\end{table*}

\begin{table*}
\begin{tabular}{|cccc|cc|}
\hline
Parameter & Symbol & Units & Prior & \multicolumn{2}{c|}{Posterior} \\
& & & & Figure \ref{fig:optfits_tfree} & Figure \ref{fig:optfits} \\
\hline
Ejecta mass & $M_{\rm ej}$ & $M_\odot$ & L[10$^{-1}$, 10$^2$] & 0.40 $^{+0.00}_{-0.01}$ & 0.26 $\pm$ 0.00 \\
Explosion energy & $E_{\rm SN}$ & erg & U[$5 \times 10^{50}$, $2 \times 10^{51}$] & L($50.72 ^{+0.03}_{-0.02}$) & L(${50.70 \pm 0.00}$) \\
Nickel fraction & $f_{\rm Ni}$ &  & L[10$^{-4}$, 1] & L($-0.62 \pm 0.03$)& L(${-1.26 \pm 0.02}$) \\
Opacity & $\kappa$ & cm$^2$ g$^{-1}$ & U[0.05, 0.2] & 0.05 $\pm$ 0.00 & 0.05 $\pm$ 0.00 \\
Non-thermal opacity & $\kappa_\gamma$ & cm$^2$ g$^{-1}$ &  L[10$^{-4}$, 10$^4$] & 0.012 $\pm 0.00$& 0.001 $\pm$ 0.000 \\
Initial magnetar luminosity & $L_0$ & erg s$^{-1}$ & L[10$^{40}$, 10$^{50}$] & L($46.74 ^{+0.02}_{-0.04}$)& L(${45.73 \pm 0.01}$)\\
Spin-down timescale & $t_{\rm SD}$ & s & L[10$^2$, 10$^8$] & L($4.74 \pm 0.03$)& L(${5.71 \pm 0.02}$)\\
Braking Index & $n$ & & U[1.5, 10] & 4.06 $^{+0.09}_{-0.10}$& 9.80 $^{+0.15}_{-0.30}$ \\
Plateau temperature & $T_{\rm floor}$ & K &  L[10$^3$, 10$^5$] (Figure \ref{fig:optfits_tfree}) & 47900 $\pm$ 200 & \\
& & & L[10$^3$, 10$^4$] (Figure \ref{fig:optfits}) && 9999 $\pm$ 1 \\
\hline
\end{tabular}
\caption{Same as Table \ref{tbl:sn_ni}, but for the magnetar-powered supernova model.}
\label{tbl:sn_mag}
\end{table*}

\begin{table*}
\begin{tabular}{|cccc|cc|}
\hline
Parameter & Symbol & Units & Prior & \multicolumn{2}{c|}{Posterior} \\
& & & & Figure \ref{fig:optfits_tfree} & Figure \ref{fig:optfits} \\
\hline
Ejecta mass & $M_{\rm ej}$ & $M_\odot$ & L[10$^{-4}$, 10$^2$] & $0.0001 \pm 0.0000$ & $1.47 \pm 0.00$\\
Explosion energy & $E_{\rm SN}$ & erg & L[$10^{48}$, $10^{52}$] & L($50.46 \pm 0.00$) & L($52.00 \pm 0.00$)\\
Nickel fraction & $f_{\rm Ni}$ &  & L[10$^{-3}$, 1] & L(-1.50$^{+1.04}_{-1.03}$) & 0.99$^{+0.01}_{-0.02}$ \\
Opacity & $\kappa$ & cm$^2$ g$^{-1}$ & U[0.05, 2] & $1.99 \pm 0.00$ & $0.05 \pm 0.00$ \\
Non-thermal opacity & $\kappa_\gamma$ & cm$^2$ g$^{-1}$ &  L[10$^{-4}$, 10$^4$] & L(-0.01$^{+2.72}_{-2.75}$) & $0.01 \pm 0.00$\\
CSM mass & $M_{\rm CSM}$ & $M_\odot$ & L[10$^{-4}$, 10$^2$] & $0.46 \pm 0.00$ & $1.28 \pm 0.00$ \\
CSM density profile index & $\eta$ &  & U[0, 2] & $0.97 \pm 0.01$ & $2.00 \pm 0.00$\\
CSM density & $\rho$ & cm$^{-3}$ & L[10$^{-15}$, 10$^{-12}$] & L($-14.64 \pm 0.01$) & L($-12.76 \pm 0.00$) \\
CSM shell radius & $r_0$ & AU & U[50, 700] & $50.0 \pm 0.1$ & $63.0 \pm 0.1$\\
Plateau temperature & $T_{\rm floor}$ & K &  L[10$^3$, 10$^5$] (Figure \ref{fig:optfits_tfree}) & $99972^{+21}_{-44}$ & \\
& & & L[10$^3$, 10$^4$] (Figure \ref{fig:optfits}) && $10000^{+0}_{-1}$\\
\hline
\end{tabular}
\caption{Same as Table \ref{tbl:sn_ni}, but for the CSM-powered supernova model.}
\label{tbl:sn_csm}
\end{table*}

\begin{table*}
\begin{tabular}{|cccc|cc|}
\hline
Parameter & Symbol & Units & Prior & \multicolumn{2}{c|}{Posterior} \\
& & & & Figure \ref{fig:optfits_tfree} & Figure \ref{fig:optfits} \\
\hline
Ejecta mass & $M_{\rm ej}$ & $M_\odot$ & L[10$^{-4}$, 10$^2$] & L($-2.7^{+1.2}_{-0.9}$) & L($-3.2^{+0.9}_{-0.5}$) \\
Ejecta velocity & $v_{\rm ej}$ & km s$^{-1}$ & L[10$^3$, 10$^5$] & L($4.62 \pm 0.00$) & L($4.78 \pm 0.00$)\\
Nickel fraction & $f_{\rm Ni}$ &  & L[10$^{-3}$, 1] & L($-1.92^{+1.07}_{-0.75}$) & L($-2.23 ^{+0.85}_{-0.54}$)\\
Opacity & $\kappa$ & cm$^2$ g$^{-1}$ & U[0.05, 2] & $1.03 \pm 0.64$ & $1.03 \pm 0.65$\\
Non-thermal opacity & $\kappa_\gamma$ & cm$^2$ g$^{-1}$ &  L[10$^{-4}$, 10$^4$] &$0.97^{+427.33}_{-0.97}$& $0.69^{+392.57}_{-0.69}$\\
Fallback luminosity & $L_1$ & erg s$^{-1}$ & L[10$^{51}$, 10$^{57}$] & L($56.03^{+0.00}_{-0.01}$) & L($53.53 \pm 0.00$)\\
Fallback transition time & $t_{\rm tr}$ & days & L[10$^{-4}$, 10$^2$] & $2.36 ^{+0.04}_{-0.05}$ & $2.16 \pm 0.03$\\
Plateau temperature & $T_{\rm floor}$ & K &  L[10$^3$, 10$^5$] (Figure \ref{fig:optfits_tfree}) & $99700^{+200}_{-500}$& \\
& & & L[10$^3$, 10$^4$] (Figure \ref{fig:optfits}) && $2230 \pm 10$\\
\hline
\end{tabular}
\caption{Same as Table \ref{tbl:sn_ni}, but for the fallback-powered supernova model.}
\label{tbl:sn_fb}
\end{table*}

\begin{table*}
\begin{tabular}{|cccc|cc|}
\hline
Parameter & Symbol & Units & Prior & \multicolumn{2}{c|}{Posterior} \\
& & & & Figure \ref{fig:optfits_tfree} & Figure \ref{fig:optfits} \\
\hline
Ejecta mass & $M_{\rm ej}$ & $M_\odot$ & L[10$^{-4}$, 10$^2$] & $12.5 \pm 0.4 $& $91.3^{+1.1}_{-0.8}$\\
Shock velocity & $v_{\rm sh}$ & km s$^{-1}$ & L[10$^3$, $5 \times 10^4$] & L($4.59 \pm 0.00$)& L($4.61 \pm 0.00$)\\
Nickel fraction & $f_{\rm Ni}$ &  & L[10$^{-3}$, 1] & 1.00 $\pm$ 0.00 & 0.001 $\pm$ 0.000 \\
Opacity & $\kappa$ & cm$^2$ g$^{-1}$ & U[0.01, 1] &$0.01 \pm 0.00$ & $0.01 \pm 0.00$\\
Non-thermal opacity & $\kappa_\gamma$ & cm$^2$ g$^{-1}$ &  L[10$^{-4}$, 10$^4$] & L($-2.88 \pm 0.03$) & L($-2.78 \pm 0.01$)\\
Envelope mass & $M_{\rm env}$ & $M_\odot$ & L[10$^{-3}$, 5] &$1.45 \pm 0.03$& $0.38 \pm 0.00$\\
Envelope structure coefficient  & $f_{\rho}$ & & U[0.5, 3] &$2.88^{+0.09}_{-0.18}$& $1.62^{+0.01}_{-0.02}$\\
Stellar radius & $R_*$ & 10$^{13}$ cm & L[10$^{-3}$, 10] &$5.62 \pm 0.18$& $3.88 \pm 0.01$\\
Plateau temperature & $T_{\rm floor}$ & K &  L[10$^3$, 10$^5$] (Figure \ref{fig:optfits_tfree}) &$31500 \pm 300$& \\
& & & L[10$^3$, 10$^4$] (Figure \ref{fig:optfits}) && $9988^{+2}_{-2}$\\
\hline
\end{tabular}
\caption{Same as Table \ref{tbl:sn_ni}, but for the SN + envelope shock cooling model.}
\label{tbl:sc_snenv}
\end{table*}

\begin{table*}
\begin{tabular}{|cccc|cc|}
\hline
Parameter & Symbol & Units & Prior & \multicolumn{2}{c|}{Posterior} \\
& & & & Figure \ref{fig:optfits_tfree} & Figure \ref{fig:optfits} \\
\hline
Ejecta mass & $M_{\rm ej}$ & $M_\odot$ & L[10$^{-4}$, 10$^2$] &$9.11^{+0.00}_{-0.01}$& $1.43^{+0.03}_{-0.05}$\\
Ejecta velocity & $v_{\rm ej}$ & km s$^{-1}$ & L[10$^3$, $10^5$] & L($4.92 \pm 0.00$)& L($4.66 \pm 0.00$)\\
Nickel fraction & $f_{\rm Ni}$ &  & L[10$^{-3}$, 1] & 1.00 $\pm$ 0.00 & 0.97 $\pm$ 0.01 \\
Opacity & $\kappa$ & cm$^2$ g$^{-1}$ & U[0.05, 2] &$0.05 \pm 0.00$ & $0.32 \pm 0.00$\\
Non-thermal opacity & $\kappa_\gamma$ & cm$^2$ g$^{-1}$ &  L[10$^{-4}$, 10$^4$] &$0.01 \pm 0.00$ & $0.04 \pm 0.00$\\
CSM mass & $M_{\rm CSM}$ & $M_\odot$ & L[10$^{-2}$, 10$^3$] &L($0.10 \pm 0.00$)& L($-0.06 \pm 0.01$)\\
CSM radius & $R_{\rm CSM}$ & 10$^{13}$ cm & L[10$^{-3}$, 10] & 8.86 $\pm$ 0.01 & 10.00 $\pm$ 0.00 \\
CSM energy & $E_{\rm CSM}$ & erg & L[10$^{40}$, 10$^{52}$] & L($51.98 \pm 0.00$)& L($51.96 \pm 0.00$)\\
Inner density index & $\delta$ & & U[1, 1.5] & $1.2 \pm 0.00$& $1.12^{+0.06}_{-0.05}$\\
Outer density index & $n$ & & U[8, 12] & $10.86 \pm 0.01$& $8.78 \pm 0.01$\\
Plateau temperature & $T_{\rm floor}$ & K &  L[10$^3$, 10$^5$] (Figure \ref{fig:optfits_tfree}) &$18492^{+4}_{-5}$& \\
& & & L[10$^3$, 10$^4$] (Figure \ref{fig:optfits}) && $9988^{+7}_{-9}$\\
\hline
\end{tabular}
\caption{Same as Table \ref{tbl:sn_ni}, but for the SN + CSM shock cooling model.}
\label{tbl:sc_sncsm}
\end{table*}

\begin{table*}
\begin{tabular}{|cccc|cc|}
\hline
Parameter & Symbol & Units & Prior & \multicolumn{2}{c|}{Posterior} \\
& & & & Figure \ref{fig:optfits_tfree} & Figure \ref{fig:optfits} \\
\hline
Ejecta mass & $M_{\rm ej}$ & $M_\odot$ & L[10$^{-1}$, 10$^2$] & $9.72^{+0.09}_{-0.20}$& 2.31$^{+0.02}_{-0.03}$ \\
Ejecta velocity & $v_{\rm ej}$ & km s$^{-1}$ & L[10$^3$, $10^5$] & L($5.00 \pm 0.00$)& L(4.37 $\pm$ 0.00)  \\
Nickel fraction & $f_{\rm Ni}$ &  & L[10$^{-3}$, 1] & $1.00 \pm 0.00$ & 1.00 $\pm$ 0.00 \\
Opacity & $\kappa$ & cm$^2$ g$^{-1}$ & U[0.05, 2] &$0.05 \pm 0.00$& 0.05 $\pm$ 0.00 \\
Non-thermal opacity & $\kappa_\gamma$ & cm$^2$ g$^{-1}$ &  L[10$^{-4}$, 10$^4$] & $0.01 \pm 0.00 $& 0.0038 $\pm$ 0.0001 \\
Cocoon mass & $M_{\rm coc}$ & $M_\odot$ & U[0.1, 2] &$1.60 \pm 0.04$ & 1.52 $\pm$ 0.02 \\
Cocoon velocity & $v_{\rm coc}$ & $c$ & U[0.01, 0.3] &$0.15 \pm 0.00$ &0.16 $\pm$ 0.00  \\
Cocoon density index & $\eta$ & & U[1, 5] &$4.97 \pm 0.01$ & 4.96 $\pm$ 0.03 \\
Shock breakout time & $t_{\rm SBO}$ & s & U[0.1, 100] &$98.5^{+0.6}_{-0.7}$ & 99.9 $\pm$ 0.01 \\
Shocked fraction of cocoon & $f_{\rm sh}$ & & U[0.01, 1] & $ 0.65 \pm 0.02$& 0.90 $\pm$ 0.01 \\
Cocoon opening angle & $\cos(\theta_{\rm coc})$ & & U[0.866, 1] &$0.87 \pm 0.00$& 0.866 $\pm$ 0.000 \\
Plateau temperature & $T_{\rm floor}$ & K &  L[10$^3$, 10$^5$] (Figure \ref{fig:optfits_tfree}) & $19980^{+90}_{-110}$& \\
& & & L[10$^3$, 10$^4$] (Figure \ref{fig:optfits}) && 8930 $\pm$ 10 \\
\hline
\end{tabular}
\caption{Same as Table \ref{tbl:sn_ni}, but for the cocoon + envelope shock cooling model.}
\label{tbl:sc_cocenv}
\end{table*}

\begin{table*}
\begin{tabular}{|cccc|cc|}
\hline
Parameter & Symbol & Units & Prior & \multicolumn{2}{c|}{Posterior} \\
& & & & Figure \ref{fig:optfits_tfree} & Figure \ref{fig:optfits} \\
\hline
Ejecta mass & $M_{\rm ej}$ & $M_\odot$ & L[10$^{-1}$, 10$^2$] & 14.3 $\pm$ 0.2 & 1.16$^{+0.00}_{-0.01}$ \\
Ejecta velocity & $v_{\rm ej}$ & km s$^{-1}$ & L[10$^3$, $10^5$] & 99100 $\pm$ 400 & 4120 $\pm$ 20 \\
Nickel fraction & $f_{\rm Ni}$ &  & L[10$^{-3}$, 1] & 1.00 $\pm$ 0.00 & 0.91 $\pm$ 0.00 \\
Opacity & $\kappa$ & cm$^2$ g$^{-1}$ & U[0.05, 2] & 0.050 $\pm$ 0.000 & 0.050 $\pm$ 0.000 \\
Non-thermal opacity & $\kappa_\gamma$ & cm$^2$ g$^{-1}$ &  L[10$^{-4}$, 10$^4$] & 0.0046$^{+0.0002}_{-0.0001}$ & 0.0005 $\pm$ 0.0000 \\
Engine energy & $E_{\rm eng}$ & erg & L[10$^{48}$, $10^{53}$] & L(51.62 $\pm$ 0.01)  & L(51.84 $\pm$ 0.00) \\
Engine timescale & $t_{\rm eng}$ & s & L[1, $10^{4}$] & 120$^{+20}_{-10}$ & 410$^{+50}_{-20}$ \\
Cocoon opening angle & $\theta_{\rm coc}$ & rad & U[0.01, 0.1] & 0.011 $\pm$ 0.001 & 0.070 $\pm$ 0.001 \\
CSM mass & $M_{\rm CSM}$ & $M_\odot$ & L[10$^{-2}$, 10] & 0.94 $\pm$ 0.01 & 1.413 $\pm$ 0.001 \\
CSM radius & $R_{\rm CSM}$ & 10$^{13}$ cm & L[10$^{-2}$, 10$^2$] & 10.3 $\pm$ 0.2 & 12.0 $\pm$ 0.0 \\
Plateau temperature & $T_{\rm floor}$ & K &  L[10$^3$, 10$^5$] (Figure \ref{fig:optfits_tfree}) & 19960$^{+70}_{-50}$ & \\
& & & L[10$^3$, 10$^4$] (Figure \ref{fig:optfits}) && 9930 $\pm$ 10 \\
\hline
\end{tabular}
\caption{Same as Table \ref{tbl:sn_ni}, but for the cocoon + CSM shock cooling model.}
\label{tbl:sc_coccsm}
\end{table*}

\begin{table*}
\begin{tabular}{|cccc|c|}
\hline
Parameter & Symbol & Units & Prior & Posterior \\
\hline
Black hole mass & $M_{\rm BH}$ & 10$^6$ $M_\odot$ & L[10$^{-6}$, 1] & 1.00 $\pm$ 0.00 \\
Stellar mass & $M_*$ &  $M_\odot$ & L[0.05, 10] & 9.99$^{+0.01}_{-0.02}$ \\
Viscous timescale & $t_{\rm visc}$ & days & L[10$^{-3}$, 10$^2$] & 4.23 $\pm$ 0.07 \\
Scaled impact parameter & $b$ &  & U[0, 2] & 0.32 $\pm$ 0.00 \\
Efficiency & $\eta$ &  & L[10$^{-4}$, 0.4] & L(-3.00 $\pm$ 0.00) \\
Photosphere power-law constant & $R_{\rm ph,0}$ & cm & L[10$^{-4}$, 10$^4$] & 53.4 $\pm$ 0.9 \\
Photosphere power-law exponent & $l$ & & U[0, 2] & 2.00 $\pm$ 0.00 \\
\hline
\end{tabular}
\caption{Same as Table \ref{tbl:sn_ni}, but for the fallback TDE model.}
\label{tbl:tde_mosfit}
\end{table*}

\begin{table*}
\begin{tabular}{|cccc|c|}
\hline
Parameter & Symbol & Units & Prior & Posterior \\
\hline
Black hole mass & $M_{\rm BH}$ & 10$^6$ $M_\odot$ & L[10$^{-6}$, 1] & 0.10$^{+0.16}_{-0.06}$  \\
Stellar mass & $M_*$ &  $M_\odot$ & U[0.05, 10] & 5.96$^{+2.46}_{-2.96}$ \\
Gaussian peak time & $t_{\rm peak}$ & day & U[0, 60] & 45.86$^{+8.60}_{-10.38}$ \\
Gaussian sharpness & $\sigma_t$ & day & U[0, 60] & 0.44$^{+0.35}_{-0.28}$ \\
\hline
\end{tabular}
\caption{Same as Table \ref{tbl:sn_ni}, but for the stream-stream TDE model.}
\label{tbl:tde_ssc}
\end{table*}

\begin{table*}
\begin{tabular}{|cccc|c|}
\hline
Parameter & Symbol & Units & Prior & Posterior \\
\hline
Black hole mass & $M_{\rm BH}$ & $M_\odot$ & L[1, 10$^6$] & 296$^{+1}_{-2}$\\
Stellar mass & $M_*$ &  $M_\odot$ & L[0.05, 10] & 10.00 $\pm$ 0.00 \\
Impact parameter & $\beta$ & & U[1, 100] & 86 $\pm$ 4 \\
Viscosity parameter & $\alpha$ &  & L[0.1, 1] & 0.50$^{+0.33}_{-0.30}$ \\
Feedback efficiency & $\eta$ & & L[10$^{-4}$, 0.1] & L(-1.85$^{+0.40}_{-0.42}$) \\
Gaussian peak time & $t_{\rm peak}$ & day & U[5, 60] & 1.57 $\pm$ 0.00 \\
Gaussian sharpness & $\sigma_t$ & day & U[5, 60] & 0.037 $\pm$ 0.000 \\
\hline
\end{tabular}
\caption{Same as Table \ref{tbl:sn_ni}, but for the cooling envelope TDE model.}
\label{tbl:tde_ce}
\end{table*}

\begin{table*}
\begin{tabular}{|cccc|cc|}
\hline
Parameter & Symbol & Units & Prior & \multicolumn{2}{c|}{Posterior} \\
& & & & Figure \ref{fig:optfits_tfree} & Figure \ref{fig:optfits} \\
\hline
Stellar mass & $M_*$ & $M_\odot$ & U[5, 50] & 46$^{+3}_{-4}$ & 50.0$^{+0.0}_{-0.1}$\\
Black hole mass & $M_{\rm BH}$ & $M_\odot$ & U[5, 50] & 6.1$^{+1.2}_{-0.8}$ & 50.0$^{+0.0}_{-0.1}$ \\
CSM mass & $M_{\rm CSM}$ & $M_\odot$ & U[0.1, 5] & 1.40$^{+0.25}_{-0.15}$ & 0.40 $\pm$ 0.04 \\
Fast component mass & $M_{\rm fast}$ & $M_\odot$ & U[0.01, 1] & 1.00 $\pm$ 0.00 & 0.088 $\pm$ 0.001 \\
Fast component velocity & $v_{\rm fast}$ & $c$ & U[0.05, 0.3] & 0.21 $\pm$ 0.00 & 0.094 $\pm$ 0.00 \\
Slow component velocity & $v_{\rm slow}$ & km s$^{-1}$ & L[10$^3$, 10$^4$] & 6300 $\pm$ 600 & 4000 $\pm$ 300 \\
Viscosity parameter & $\alpha$ &  & L[0.1, 1] & 0.11 $\pm$ 0.01 & 0.210 $\pm$ 0.001 \\
Feedback efficiency & $\eta$ & & L[10$^{-4}$, 0.1] & 0.069 $\pm$ 0.09 & 0.10 $\pm$ 0.00 \\
Plateau temperature & $T_{\rm floor}$ & K &  L[10$^3$, 10$^5$] (Figure \ref{fig:optfits_tfree}) & 28100$^{+200}_{-100}$ & \\
& & & L[10$^3$, 10$^4$] (Figure \ref{fig:optfits}) & & 9999 $\pm$ 1 \\
\hline
\end{tabular}
\caption{Same as Table \ref{tbl:sn_ni}, but for the WR + BH merger model.}
\label{tbl:wr_bh}
\end{table*}

\begin{table*}
\begin{tabular}{|cccc|cc|}
\hline
Parameter & Symbol & Units & Prior & \multicolumn{2}{c|}{Posterior} \\
& & & & Radio & X-ray \\
\hline
Observer angle & $\theta_{\rm ob}$ & rad & Sine[0, $\pi$/2] & 0.187$^{+0.000}_{-0.001}$ & 0.28$^{+0.07}_{-0.06}$\\
Jet opening angle & $\theta_c$ & rad & U[0.01, 0.1] & 0.028 $\pm$ 0.001 & 0.075$^{+0.017}_{-0.020}$ \\
Isotropic equivalent jet energy & $E_{\rm jet}$ & erg & L[10$^{44}$, 10$^{54}$] & L(53.0$^{+0.3}_{-0.4}$) & L(52.4$^{+0.9}_{-0.6}$) \\
ISM density & $n_0$ & cm$^{-3}$ & L[10$^{-5}$, 10$^2$] & L(-2.3$^{+0.3}_{-0.4}$) & L(1.1$^{+0.5}_{-0.9}$) \\
Electron spectral index & $p$ & & U[2, 3] & 2.13 $\pm$ 0.03 & 2.32$^{+0.31}_{-0.16}$ \\
Particle equipartition parameter & $\epsilon_e$ & & L[10$^{-5}$, 1] & L(-0.33 $\pm$ 0.20) & L(-0.69$^{+0.44}_{-0.69}$) \\
Magnetic field equipartition parameter & $\epsilon_B$ & & L[10$^{-5}$, 1] & L(-1.5 $\pm$ 1.0) & L(-1.9$^{+1.2}_{-1.4}$) \\
Initial Lorentz factor & $\Gamma_0$ & & U[100,2000] & 160$^{+250}_{-50}$ & 1000$^{+600}_{-500}$ \\
Accelerated electron fraction & $\xi_N$ & & U[0,1] & 0.79$^{+0.13}_{-0.16}$ & 0.51 $\pm$ 0.30 \\
\hline
\end{tabular}
\caption{Priors and summary statistics for the radio and X-ray top-hat jet afterglow models in Figure \ref{fig:radx_lcs}. Priors are either uniform (U) or log-uniform (L). The values shown for the posterior are the median and 1$\sigma$ uncertainties. Posterior values denoted with L are given in log-space.}
\label{tbl:tophat}
\end{table*}

\begin{table*}
\begin{tabular}{|cccc|cc|}
\hline
Parameter & Symbol & Units & Prior & \multicolumn{2}{c|}{Posterior} \\
& & & & Radio & X-ray \\
\hline
Observer angle & $\theta_{\rm ob}$ & rad & Sine[0, $\pi$/2] & 0.88 $\pm$ 0.00 & 0.35$^{+0.06}_{-0.07}$ \\
Jet core angle & $\theta_c$ & rad & U[0.01, 0.1] & 0.010 $\pm$ 0.000 & 0.056$^{+0.024}_{-0.019}$ \\
Jet edge angle & $\theta_j$ & rad & U[0.1, 0.2] & 0.10 $\pm$ 0.00 & 0.14$^{+0.04}_{-0.02}$ \\
Isotropic equivalent jet energy & $E_{\rm jet}$ & erg & L[10$^{44}$, 10$^{54}$] & L(52.97 $\pm$ 0.01) & L(52.3$^{+0.7}_{-0.4}$)\\
ISM density & $n_0$ & cm$^{-3}$ & L[10$^{-5}$, 10$^2$] & L(2.00 $\pm$ 0.00) & L(1.4$^{+0.4}_{-0.7}$) \\
Electron spectral index & $p$ & & U[2, 3] & 2.38$^{+0.01}_{-0.02}$ & 2.35$^{+0.29}_{-0.19}$\\
Particle equipartition parameter & $\epsilon_e$ & & L[10$^{-5}$, 1] & L(-0.96$^{+0.01}_{-0.02}$) & L(-0.45$^{+0.29}_{-0.46}$) \\
Magnetic field equipartition parameter & $\epsilon_B$ & & L[10$^{-5}$, 1] & L(-0.75 $\pm$ 0.02) & L(-1.9$^{+1.2}_{-1.4}$) \\
Initial Lorentz factor & $\Gamma_0$ & & U[100,2000] & 1400$^{+400}_{-500}$ & 1100$^{+500}_{-600}$ \\
Accelerated electron fraction & $\xi_N$ & & U[0,1] & 1.00$^{+0.00}_{-0.01}$ & 0.54$^{+0.29}_{-0.33}$ \\
\hline
\end{tabular}
\caption{Same as Table \ref{tbl:tophat}, but for the radio and X-ray gaussian jet afterglow models.}
\label{tbl:gaussian}
\end{table*}

\begin{table*}
\begin{tabular}{|cccc|cc|}
\hline
Parameter & Symbol & Units & Prior & \multicolumn{2}{c|}{Posterior} \\
& & & & Radio & X-ray \\
\hline
Shock velocity & $v_{\rm shock}$ & km s$^{-1}$ & U[10$^3$, $3 \times 10^4$] & 29700$^{+200}_{-300}$  & 18000$^{+8000}_{-9000}$ \\
CSM density at 10$^{15}$ cm & $A$ & cm$^{-3}$ & L[10$^{-5}$, 10$^{15}$] & L(9.63 $\pm$ 0.06) & L(9.1 $\pm$ 1.3)\\
CSM density index & $s$ & & U[0,4] & 2.86 $\pm$ 0.01 & 2.06 $\pm$ 0.11 \\
Electron spectral index & $p$ & & U[2, 3] & 2.34 $\pm$ 0.03 & 2.4$^{+0.4}_{-0.3}$\\
Particle equipartition parameter & $\epsilon_e$ & & L[10$^{-5}$, 1] & L(-0.07$^{+0.05}_{-0.07}$) & L(-2.5 $\pm$ 1.7) \\
Magnetic field equipartition parameter & $\epsilon_B$ & & L[10$^{-5}$, 1] & L(-4.94$^{+0.08}_{-0.04}$) & L(-2.6$^{+1.7}_{-1.5}$)\\
\hline
\end{tabular}
\caption{Same as Table \ref{tbl:tophat}, but for the radio and X-ray power-law synchrotron blast wave models.}
\label{tbl:synch}
\end{table*}

\begin{table*}
\begin{tabular}{|cccc|cc|}
\hline
Parameter & Symbol & Units & Prior & \multicolumn{2}{c|}{Posterior} \\
& & & & Radio & X-ray \\
\hline
Shock velocity & $\beta\Gamma$ &  & U[0.1, 10] & 0.13 $\pm$ 0.01 & 1.5 $\pm$ 0.7 \\
ISM density & $n_{\rm ISM}$ & cm$^{-3}$ & U[0, 10] & 3.5$^{+3.7}_{-2.6}$ & 4.4$^{+3.2}_{-2.9}$ \\
Mass loss parameter & $\dot{M}v_{\rm wind}$ & $M_\odot$ s km$^{-1}$ year$^{-1}$ & L[10$^{-10}$, 1] & L(-5.8$^{+0.3}_{-0.5}$) & L(-5.7$^{+0.6}_{-0.7}$) \\
Electron spectral index & $p$ & & U[2, 3] & 2.02 $\pm$ 0.01 & 2.94$^{+0.04}_{-0.09}$ \\
Particle equipartition parameter & $\epsilon_e$ & & L[10$^{-5}$, 1] & L(-1.7$^{+0.5}_{-0.7}$) & L(-3.0$^{+1.8}_{-1.3}$) \\
Magnetic field equipartition parameter & $\epsilon_B$ & & L[10$^{-5}$, 1] & L(-4.4 $\pm$ 0.4) & L(-2.6$^{+1.7}_{-1.6}$) \\
Accelerated electron fraction & $\xi_N$ & & U[0,1] & 0.64$^{+0.20}_{-0.26}$ & 0.54$^{+0.28}_{-0.29}$ \\
\hline
\end{tabular}
\caption{Same as Table \ref{tbl:tophat}, but for the radio and X-ray thermal synchrotron blast wave models.}
\label{tbl:thermsynch}
\end{table*}

\begin{table*}
\begin{tabular}{|cccc|c|}
\hline
Parameter & Symbol & Units & Prior & Posterior \\
\hline
Black hole mass & $M_{\rm BH}$ & $M_\odot$ & L[1, 10$^{10}$] & L(5.6$^{+1.2}_{-0.7}$) \\
Black hole spin & $a$ & & U[-1, 1] & -0.48$^{+0.50}_{-0.32}$ \\
Disc mass & $M_{\rm disc}$ & $M_\odot$ & L[10$^{-3}$, 10] & L(-2.02$^{+1.94}_{-0.61}$) \\
Initial radius of disc & $R_0$ & $R_{\rm grav}$ & U[10, 100] & 79$^{+12}_{-19}$ \\
Viscous timescale & $t_{\rm visc}$ & day & L[1, 10$^3$] & 4.2$^{+5.7}_{-2.4}$ \\
Time of ring formation & $t_{\rm form}$ & day & U[-100, 365] & 12.8$^{+9.3}_{-6.1}$ \\
Observer angle & $\theta_{\rm obs}$ & rad & Sine[0, $\pi$/2] & 0.88$^{+0.37}_{-0.33}$ \\
\hline
\end{tabular}
\caption{Same as Table \ref{tbl:tophat}, but for the X-ray accretion disc model.}
\label{tbl:fitted}
\end{table*}

\begin{table*}
\begin{tabular}{|cccc|c|}
\hline
Parameter & Symbol & Units & Prior & Posterior \\
\hline
Black hole mass & $M_{\rm BH}$ & $M_\odot$ & L[1, 10$^{10}$] & L(5.3$^{+0.4}_{-0.5}$) \\
Black hole spin & $a$ & & U[-1, 1] & -0.43$^{+0.40}_{-0.33}$ \\
Disc mass & $M_{\rm disc}$ & $M_\odot$ & L[10$^{-3}$, 10] & 0.004$^{+0.010}_{-0.002}$\\
Initial radius of disc & $R_0$ & $R_{\rm grav}$ & U[10, 100] & 43$^{+25}_{-19}$ \\
Viscous timescale & $t_{\rm visc}$ & day & L[1, 10$^3$] & 4.9$^{+6.0}_{-2.6}$ \\
Time of ring formation & $t_{\rm form}$ & day & U[-100, 365] & 11.1$^{+4.9}_{-3.4}$ \\
Observer angle & $\theta_{\rm ob}$ & rad & Sine[0, $\pi$/2] & 1.25$^{+0.19}_{-0.32}$ \\
Isotropic equivalent jet energy & $E_{\rm jet}$ & erg & L[10$^{46}$, 10$^{53}$] &  L(50.3$^{+1.5}_{-2.0}$) \\
ISM density & $n_0$ & cm$^{-3}$ & L[10$^{-5}$, 10$^2$] & L(-1.1$^{+1.8}_{-1.9}$) \\
Particle equipartition parameter & $\epsilon_e$ & & L[10$^{-5}$, 1] & L(-2.2$^{+1.2}_{-1.5}$) \\
Magnetic field equipartition parameter & $\epsilon_B$ &  &L[10$^{-5}$, 1] & L(-2.8 $\pm$ 1.3)  \\
\hline
Jet opening angle & $\theta_c$ & rad & 0.1 & \\
Electron spectral index & $p$ & & 2.5 & \\
Initial Lorentz factor & $\Gamma_0$ & & 100 & \\
Accelerated electron fraction & $\xi_N$ & & 1 & \\
\hline
\end{tabular}
\caption{Same as Table \ref{tbl:tophat}, but for the X-ray top-hat jet afterglow + accretion disc model.}
\label{tbl:tophatfitted}
\end{table*}

\begin{table*}
\begin{tabular}{|cccc|c|}
\hline
Parameter & Symbol & Units & Prior & Posterior \\
\hline
Shock velocity & $v_{\rm shock}$ & km s$^{-1}$ & U[10$^3$, $3 \times 10^4$] & 29996$^{+3}_{-6}$\\
CSM density at 10$^{15}$ cm & $A$ & cm$^{-3}$ & L[10$^{-5}$, 10$^{15}$] & L(10.02$^{+0.03}_{-0.02}$) \\
CSM density index & $s$ & & U[0,4] & 2.67 $\pm$ 0.01 \\
Electron spectral index & $p$ & & U[2, 3] & 2.85$^{+0.02}_{-0.01}$\\
Particle equipartition parameter & $\epsilon_e$ & & L[10$^{-5}$, 1] & L(-0.0010$^{+0.0007}_{-0.0018}$) \\
Magnetic field equipartition parameter & $\epsilon_B$ & & L[10$^{-5}$, 1] & L(-4.99 $\pm$ 0.00) \\
\hline
\end{tabular}
\caption{Priors and summary statistics for the joint radio and X-ray power-law synchrotron blast wave model in Figure \ref{fig:radx_combined_lcs}. Priors are either uniform (U) or log-uniform (L). The values shown for the posterior are the median and 1$\sigma$ uncertainties. Posterior values denoted with L are given in log-space.}
\label{tbl:synch_joint}
\end{table*}

\begin{table*}
\begin{tabular}{|cccc|c|}
\hline
Parameter & Symbol & Units & Prior & Posterior \\
\hline
Observer angle & $\theta_{\rm obs}$ & rad & Sine[0, $\pi$/2] & 0.610$^{+0.049}_{-0.005}$ \\
Isotropic equivalent jet energy & $E_{\rm jet}$ & erg & L[10$^{46}$, 10$^{53}$] & L(50.78$^{+0.05}_{-0.04}$) \\
ISM density & $n_0$ & cm$^{-3}$ & L[10$^{-5}$, 10$^2$] & L(-0.62$^{+0.16}_{-0.05}$) \\
Particle equipartition parameter (afterglow) & $\epsilon_{\rm e, GRB}$ & & L[10$^{-5}$, 1] & L(-0.028$^{+0.16}_{-0.20}$) \\
Magnetic field equipartition parameter (afterglow) & $\epsilon_{\rm B, GRB}$ & & L[10$^{-5}$, 1] & L(-0.22 $\pm$ 0.11)\\
Shock velocity & $v_{\rm shock}$ & km s$^{-1}$ & U[10$^3$, $3 \times 10^4$] & 25000$^{+3000}_{-4000}$\\
CSM density at 10$^{15}$ cm & $A$ & cm$^{-3}$ & L[10$^{-5}$, 10$^{15}$] & L(8.6$^{+0.6}_{-0.9}$) \\
CSM density index & $s$ & & U[0,4] & 2.34 $\pm$ 0.00 \\
Particle equipartition parameter (blast wave) & $\epsilon_{\rm e, ej}$ & & L[10$^{-5}$, 1] & L(-1.19$^{+0.67}_{-0.90}$) \\
Magnetic field equipartition parameter (blast wave) & $\epsilon_{\rm B, ej}$ & & L[10$^{-5}$, 1] & L(-2.5$^{+1.3}_{-0.9}$) \\
\hline
Jet core angle & $\theta_c$ & rad & 0.1 & \\
Jet edge angle & $\theta_j$ & rad & 0.2 & \\
Electron spectral index (afterglow) & $p_{\rm GRB}$ & & 2.5 & \\
Initial Lorentz factor & $\Gamma_0$ & & 100 & \\
Accelerated electron fraction & $\xi_N$ & & 1 & \\
Electron spectral index (blast wave) & $p_{\rm ej}$ & & 2.5 & \\
\hline
\end{tabular}
\caption{Same as Table \ref{tbl:synch_joint}, but for the power-law synchrotron blast wave + gaussian jet afterglow model.}
\label{tbl:synchgauss_joint}
\end{table*}

\section{The Photosphere of a Homologous Expanding Medium} \label{sec:photo}

One argument against interpretations involving SNe, jets, or other expanding media is the behaviour of the photosphere of AT2024wpp.  The photosphere peaks at $\sim$ 5 days, but the spectra suggest that ejecta remains optically thick until $\gtrsim$ 40 days \citep{Perley2026}.  The timescale between when the photospheric radius peaks and when the ejecta becomes optically thin can be derived using a few simple assumptions and compared to the observed data.

Assume that the ejecta is expanding homologously with a power law density profile $\rho \propto v^{-\alpha}$, where $\alpha \geq 0$.  The total mass of the ejecta is $M_{\rm ej}$ and the inner and outer edges of the ejecta are traveling at $v_{\rm in}$ and $v_{\rm out}$ respectively.  The opacity $\kappa$ is assumed to be constant and the ejecta is assumed to be spherical.  The initial size of the object is initially assumed to be small, although we show the effect of relaxing that assumption after.

The photospheric radius $R_{\rm phot}$ can be derived by finding the location where the integrated optical depth from the edge of the ejecta becomes $\sim$ 2/3 \citep{Arnett1980, Arnett1989}, i.e.

\begin{equation}
    2/3 = \int_{R_{\rm phot}}^{v_{\rm out}t} \rho(r,t)\kappa dr .
    \label{eqn:rphot_1}
\end{equation}
Instead of continuing the derivation for arbitrary $\alpha$, we instead focus on the two extreme cases, $\alpha = 0$ and $\alpha \rightarrow \infty$.  

For $\alpha = 0$, the density is

\begin{equation}
    \rho = \frac{3M_{\rm ej}}{4\pi (v_{\rm out}^3 - v_{\rm in}^3)t^3} .
    \label{eqn:density}
\end{equation}
Substituting this into Equation \ref{eqn:rphot_1} gives

\begin{align}
    2/3 =& \int_{R_{\rm phot}}^{v_{\rm out}t} \frac{3M_{\rm ej}\kappa}{4\pi (v_{\rm out}^3 - v_{\rm in}^3)t^3}dr , \\
    =& \frac{3M_{\rm ej}\kappa}{4\pi (v_{\rm out}^3 - v_{\rm in}^3)t^3}(v_{\rm out}t - R_{\rm phot}) ,
\end{align}
so

\begin{equation}
    R_{\rm phot} = v_{\rm out}t - \frac{8\pi (v_{\rm out}^3 - v_{\rm in}^3)t^3}{9M_{\rm ej}\kappa} .
    \label{eqn:rphot}
\end{equation}
The timescale for the photosphere to reach a peak is

\begin{equation}
    t_{\rm phot, peak} = \sqrt{{\frac{3M_{\rm ej} \kappa v_{\rm out}}{8\pi (v_{\rm out}^3 - v_{\rm in}^3)}}} ,
    \label{eqn:trph_peak}
\end{equation}
and the timescale when the entire ejecta becomes optically thin ($R_{\rm phot} = v_{\rm in}t$) and enters the nebular phase is

\begin{equation}
    t_{\rm neb} = \sqrt{{\frac{9M_{\rm ej} \kappa (v_{\rm out} - v_{\rm in})}{8\pi (v_{\rm out}^3 - v_{\rm in}^3)}}}.
    \label{eqn:t_neb}
\end{equation}
The ratio between these two is

\begin{equation}
    \frac{t_{\rm neb}}{t_{\rm phot, peak}} = \sqrt{\frac{3(v_{\rm out} - v_{\rm in})}{v_{\rm out}}} \leq \sqrt{3}.
\end{equation}
For $\alpha \rightarrow \infty$, the ejecta expands in a thin shell.  $R_{\rm phot}$ will increases linearly until the ejecta becomes optically thin, so the ratio $t_{\rm neb}/t_{\rm phot, peak} \rightarrow 1$ for this case.  Intermediate values of $\alpha$ should show $t_{\rm neb}/t_{\rm phot, peak}$ between 1 and $\sqrt{3}$, so the highest this ratio can go within the assumptions we have made is $\sqrt{3}$.  Values higher than this indicate that the ejecta is likely highly non-spherical or not expanding homologously, either decelerating strongly, remaining stationary, or having mass continuously injected into the ejecta.

However, if the ejecta has an initial radius of $R_0$, then time $t$ in Equation \ref{eqn:rphot} gets replaced with $t + R_0/v_{\rm out}$, and both $t_{\rm phot, peak}$ and $t_{\rm neb}$ become smaller by $R_0/v_{\rm out}$, thus, the ratio between them becomes

\begin{equation}
    \frac{t_{\rm neb}}{t_{\rm phot, peak}} = \frac{t_{\rm neb(R_0 = 0)}  - R_0/v_{\rm out}}{t_{\rm phot, peak(R_0 = 0)} - R_0/v_{\rm out}} > \frac{t_{\rm neb(R_0 = 0)}}{t_{\rm phot, peak(R_0 = 0)}} .
\end{equation}
Thus, an extended progenitor on the order of $R_0 = v_{\rm out}t_{\rm phot, peak(R_0 = 0)}$ can have a high enough ratio to be consistent with observations.  However, for AT 2024wpp, this radius is $\gtrsim 10^{15}$ cm, which rules out any known type of progenitor star.

\bsp	
\label{lastpage}
\end{document}